\documentclass[10pt,letterpaper,twocolumn]{article}
\usepackage[T1]{fontenc}
\usepackage{cite}
\usepackage[utf8]{inputenc}
\usepackage{mathptmx,amsmath,amssymb,slashed,graphicx,booktabs,microtype}
\usepackage[margin=0.65in,columnsep=0.25in]{geometry}
\usepackage[font=small,labelfont=bf,labelsep=period]{caption}
\usepackage{xcolor,fancyhdr}
\usepackage{balance,titlesec}
\usepackage{comment}
\titleformat{\section}{\normalsize\bfseries}{\thesection.}{0.6em}{}
\titlespacing*{\section}{0pt}{11pt plus 2pt minus 1pt}{5pt plus 1pt}
\usepackage[colorlinks=false,linkcolor=black,citecolor=black,urlcolor=blue!45!black]{hyperref}
\renewcommand{\thetable}{\Roman{table}}
\newcommand{\Tr}{\mathop{\rm Tr}}
\newcommand{\SO}{\mathrm{SO}}
\newcommand{\SU}{\mathrm{SU}}
\newcommand{\U}{\mathrm{U}}
\newcommand{\Lag}{\mathcal L}
\newcommand{\bid}
{(\mathbf2,\mathbf2)_0}
\newcommand{\trip}{(\mathbf3,\mathbf2)_{1/2}}
\newcommand{\para}[1]{\par\smallskip\noindent\textit{#1 ---}}

\hypersetup{pdftitle={StronXeno: Thermal Composite Dark Matter at LZ},pdfauthor={Dario Buttazzo, Andrea Di Lecce, Roberto Franceschini, Giulio Marino, Paolo Panci},pdfsubject={Thermal composite Dark Matter and the LZ high-energy nuclear recoil}}
\begin{document}
\twocolumn[{
\begin{center}
{\Large\bfseries  Thermal Composite Dark Matter at LZ\par}
\vspace{9pt}
{Dario Buttazzo,$^{1,\ast}$ Andrea Di Lecce,$^{1,2,\dagger}$ Roberto Franceschini,$^{3,\ddagger}$\par
Giulio Marino$^{1,4,\S}$ and Paolo Panci$^{1,4,}$\textsuperscript{\P}\par}
\vspace{6pt}
{\small\itshape
$^1$INFN Sezione di Pisa, Polo Fibonacci, Largo B. Pontecorvo 3, 56127 Pisa, Italy\par
$^2$Scuola Normale Superiore, Piazza dei Cavalieri 7, 56126 Pisa, Italy\par
$^3$Universit\`a degli Studi Roma Tre and INFN, Via della Vasca Navale 84, 00146 Rome, Italy\par
$^4$Dipartimento di Fisica ``E. Fermi'', Universit\`a di Pisa, Largo Pontecorvo 3, 56127 Pisa, Italy\par}
\vspace{4pt}
\end{center}
\vspace{2pt}
\begin{quote}\small
We show that stable fermionic resonances of Composite Higgs models provide a thermal inelastic Dark Matter interpretation of the high-energy nuclear-recoil event LZ230616. Their stability follows from an exact accidental parity of the strong sector.
Annihilation channels from composite interactions link the thermal relic mass to the strong coupling, allowing masses well beyond the pure electroweak prediction. We identify three scenarios in which the neutral states form a split pseudo-Dirac pair, distinguished by the origin of the inelastic coupling to the $Z$ boson.
In the pure electroweak regime, the pair consists of the neutral components of a complex Higgsino-like doublet. Its unsuppressed inelastic $Z$ coupling requires Dark Matter masses above 20 TeV. In the pure strong regime, the suppressed $Z$ coupling of neutral states with vanishing hypercharge arises from strong interactions alone, allowing for smaller splittings and TeV-scale masses. In the third regime, which interpolates between the previous two, the neutral components of multiplets with different hypercharge are mixed. This enhances their inelastic coupling, accommodating the LZ event for masses of a few TeV and splittings around 300 keV, with mixing angles of order $10^\circ$. All three cases provide a thermal Dark Matter candidate able to explain the LZ event, consistently with the solar-capture bounds, the LZ high-energy sideband, and gamma-ray indirect searches.

\end{quote}
\vspace{7pt}
}]

\begingroup
\renewcommand{\thefootnote}{}%
\makeatletter
\long\def\@makefntext#1{\noindent #1}
\makeatother
\hypersetup{hidelinks}
\footnotetext{%
  \raggedright\footnotesize
  $\ast$ \href{mailto:dario.buttazzo@pi.infn.it}{\texttt{dario.buttazzo@pi.infn.it}}\par
  $\dagger$ \href{mailto:andrea.dilecce@sns.it}{\texttt{andrea.dilecce@sns.it}}\par
  $\ddagger$  \href{mailto:roberto.franceschini@uniroma3.it}{\texttt{roberto.franceschini@uniroma3.it}}\par
  $\S$ \href{mailto:giulio.marino@phd.unipi.it}{\texttt{giulio.marino@phd.unipi.it}}\par
\textsuperscript{\P} \href{mailto:paolo.panci@unipi.it}{\texttt{paolo.panci@unipi.it}}%
}
\endgroup

The LUX-ZEPLIN (LZ) experiment recently reported a nuclear recoil event at $E_R=248\pm23~{\rm(stat)}\pm23~{\rm(sys)}
~{\rm keV}$ in a $2.84~{\rm tonne\,yr}$ exposure
~\cite{LZ:2026axp}. Although a single event cannot establish
a Dark Matter (DM) signal, its high recoil energy motivates
endothermic inelastic scattering, $\chi_0 A\to\chi_1 A$,
between states separated by a small mass splitting
$\delta=m_{\chi_1}-m_{\chi_0}>0$. Electroweak (EW) multiplets provide a predictive and well motivated framework for thermal DM~\cite{Cirelli:2005uq,Cirelli:2007xd,DelNobile:2015bqo,Bottaro:2021snn,Bottaro:2022one,Panci:2024ewmultiplets,Cirelli:2024darkmatter,Baumgart:2017nsr}. The Higgsino offers a minimal realization of inelastic DM~\cite{Tucker-Smith:2001myb}: its neutral components form a pseudo-Dirac pair coupled by an off-diagonal $Z$ current~\cite{Fan:2026kxx,Freese:2026sga,Wu:2026nhi,Du:2026guj,Bisal:2026lz,Langhoff:2026ujr,Chatterjee:2026lz,Frolovsky:2026lz,Khan:2026susy331,Pallis:2026lz,Khan:2026gmsugra,Hor:2026lz,Freese:2026icecube,Jeong:2026lz,Wang:2026lz}. A splitting of a few hundred keV can then accommodate LZ230616. However, the large scattering rate leads to strong
solar-neutrino constraints~\cite{IceCube:2025fcu,Pospelov:2026ewn,Bose:2026ndd,Luengas:2026tlf}, while the predicted recoil spectrum can extend to higher energies, where the absence of additional events provides a complementary constraint~\cite{Rodd:2026tyn}.

Composite Higgs models \cite{Kaplan:1983fs,Kaplan:1983sm,Kaplan:1991dc, Georgi:1984af, Georgi:1984ef,Dugan:1984hq,mchm, Panico:2016,DeSimone:2012fs,Matsedonskyi:2013nsa} provide a different realization.
In Ref.~\cite{SO4neutralinos_DiLecce} we identified a new class of stable fermionic resonances in the Minimal Composite Higgs Models, where the Higgs is a pseudo Nambu Goldstone boson of the coset $\SO(5)/\SO(4)$. In this context, fermionic resonances transform under the unbroken global symmetry of the composite sector $\SU(2)_L\times\SU(2)_R\times\U(1)_X$, where hypercharge is $Y=T_R^3+X$. The DM stability is protected by an accidental $\mathbb{Z}_2$ parity embedded in the global $\U(1)_X$ symmetry group of the strong sector, rather than by an \textit{ad-hoc} discrete assumption. When decomposed under the SM gauge group, these resonances contain EW multiplets with calculable gauge interactions, while their composite origin supplies additional couplings. Crucially, strong dynamics controls thermal freeze-out, increasing the DM mass compared to the EW prediction, and generates neutral currents that are absent in pure EW multiplets. The framework remains sharply predictive: once the representation and thermal DM mass are specified, the interaction relevant for direct detection (DD) is fixed, while the neutral-state splitting is controlled by a single symmetry-breaking parameter.

This framework provides a natural stable thermal DM candidate, fixes its relic mass through a single strong coupling, and predicts its inelastic-scattering phenomenology. We identify regions of parameter space in which these candidates account for LZ230616 while satisfying constraints from the LZ high-energy sideband, solar capture, and Galactic $\gamma$-ray searches.

\medskip

\para{The composite WIMPs}
Following Ref.~\cite{SO4neutralinos_DiLecce}, we denote the composite fermionic resonances by $\psi\sim(\mathbf n,\mathbf m)_X$. We consider resonances in the two smallest $\SO(4)$ representations with the desired inelastic-scattering phenomenology, yielding substantially different predictions.

The real bidoublet $\bid$ has the field content of a Higgsino. Strong-sector annihilation allows for larger thermal masses while keeping its EW couplings responsible for the $Z$-mediated inelastic scattering. The complex $\trip$ contains two real triplets: a $\mathbf3_0$ with vanishing hypercharge and a complex triplet $\mathbf3_1$, yielding two neutral pseudo-Dirac pairs. For the Wino-like pair, inelastic scattering arises from strong interactions alone and it is therefore suppressed, whereas the hypercharged pair has a phenomenology similar to that of a Higgsino. After EW symmetry breaking (EWSB), a mixing between the neutral components of the triplets is generically expected once $\SO(5)$-breaking operators are included. 
The lightest Wino-like mass eigenstate then inherits a gauge coupling with the $Z$ boson through the mixing. An additional small splitting between its two Majorana components makes scattering off nuclei inelastic. DM is always identified with the lightest neutral state $\chi_0$.

\smallskip
We characterize the composite sector in terms of the Goldstone decay constant $f$ --- or equivalently $\xi\equiv v^2/f^2$, with
$v=246\,{\rm GeV}$ --- and a characteristic coupling of the fermionic resonances $g_\psi$, not necessarily identical to the coupling of other strong sector states. The DM resonance mass is given by $M_\chi=g_\psi f$, with the strong coupling expected to satisfy $g_\psi\gtrsim1$.

A distinctive feature of the composite resonances is
their $\SO(5)$-preserving strong interaction with the Higgs~\cite{Panico:2016}
\begin{equation}
\Lag_{\rm strong}
=-\frac{\xi}{v^2}
\left(
J_{\psi,L}^{\mu,a}J_{H,L\,\mu}^{a}
+J_{\psi,R}^{\mu,a}J_{H,R\,\mu}^{a}
\right) \ ,
\label{eq:strong}
\end{equation}
where $J_{L,R}$ are the $\SU(2)_{L,R}$ DM and Higgs currents  defined in Appendix~\ref{app:currents}.  This interaction provides additional annihilation
channels into Higgs and Goldstone bosons which determine the thermal mass, and at the same time induces a weak neutral-current coupling proportional to $\xi$.  For multiplets with vanishing hypercharge this provides a link between strong annihilation and inelastic DD, a distinctive feature of composite-Higgs DM. By contrast, a pure EW state with $Y = 0$ (such as a Wino-like triplet) has no tree-level $Z$ current. Inelastic scattering thus requires additional states or couplings, and is not intrinsically related to the dynamics responsible for DM freeze-out.

The thermal relic abundances, including
annihilations and coannihilations of the full
multiplets, are computed in
Ref.~\cite{SO4neutralinos_DiLecce}. The real bidoublet $\bid$ and the complex
$\trip$ contain
8 and 24 degrees of freedom,
respectively. Since the relevant mass splittings
are much smaller than the freeze-out temperature
$T_{\rm fo}\simeq M_\chi/25$, all components
participate in coannihilations.
In the weak-coupling limit, the thermal masses
approach $1.1\,{\rm TeV}$ for $\bid$ and
$1.6\,{\rm TeV}$ for $\trip$. The latter value is lower than the pure Wino thermal mass, because the full $\trip$ contains more internal degrees of freedom, diluting the effective annihilation rate. In the strong-annihilation regime,
$g_{\rm th}\propto\sqrt{M_\chi}$, with
$g_{\rm th}^2\simeq4.
6\,M_\chi/(10\,{\rm TeV})$
and $3.2
\,M_\chi/(10\,{\rm TeV})$ for
$\bid$ and $\trip$, respectively.
Consequently, $\xi M_\chi$ becomes
approximately constant along the thermal
trajectory. Throughout our numerical analysis,
we use the full relic-density calculation,
including the irreducible EW contributions.

\medskip
\para{Inelastic Dark Matter from composite fermions} 
Within a given EW multiplet, states with different electric charge $Q$ have their masses split by loops of gauge bosons by an amount $\Delta_{\rm EW} \equiv M_{Y}^{Q}-M_{Y}^{Q=0} \approx 167\,\text{MeV}(Q^2 + 2Q Y/c_W)$, with $c_W\equiv \cos\theta_W$~\cite{Thomas:1998wy,Nagata:2014aoa, Cheng:1998hc, Feng:1999fu, Gherghetta:1999sw, Ibe:2012sx, McKay:2017xlc}. Complex resonances receive additional mass splittings between EW multiplets with different hypercharge. For the $\trip$, the total splitting between the neutral components of the two triplets is~\cite{Yamada_2010}
\begin{equation}
\Delta_{10}\equiv M_{Y=1}^{Q=0}-M_{Y=0}^{Q=0}=
c\,\frac{\alpha_1}{4\pi}M_\chi
+\frac{\alpha_2}{2}
\left(m_W-\frac{m_Z}{c_W^2}\right),
\label{eq:Delta10-main}
\end{equation}
where $\alpha_{1,2}=g_{1,2}^2/(4\pi)$. Here, in addition to the finite IR contribution, which is approximately $-0.65\,\mathrm{GeV}$, a UV-sensitive term proportional to the DM mass arises. The coefficient $c\sim\mathcal{O}(1)$ includes the hypercharge loop contribution $4+3\log(\Lambda^2/M_\chi^2)$, together with contributions from any local operator at the cut-off $\Lambda$ that will be discussed below.
Further details are given in Appendix~\ref{app:breaking-op}.

\smallskip
The local operators contributing to the mass splittings in
Eq.~\eqref{eq:Delta10-main} can be classified according to their dimension and symmetry-breaking properties. The strong interactions in Eq.~\eqref{eq:strong} respect the full
$\SO(5)\times\U(1)_X$ symmetry and leave the resonance components
degenerate. Additional interactions may break the global symmetry
while preserving the SM gauge group and the stabilizing parity.

At dimension three, any non-universal mass term must break $\SO(4)$. The real bidoublet $\bid$ admits no such operator. For the complex $\trip$ two independent structures are allowed,
\begin{equation}
M_\chi\varepsilon_3\,
\Tr\!\left[\bar\psi\psi T_R^3\right]
+
\frac{M_\chi}{2}\slashed{\varepsilon}_3
\left(
\Tr\!\left[
\bar{\psi}\psi^c\,T_R^-
\right]
+\mathrm{h.c.}
\right).
\label{eq:HXM_dim3}
\end{equation}
The $\U(1)_X$-preserving term, parametrized by $\varepsilon_3$, splits the $Y=0$ and $Y=1$ Dirac triplets without mixing them and contributes to Eq.~\eqref{eq:Delta10-main}. The $\U(1)_X$-violating term, parametrized by $\slashed{\varepsilon}_3$, generates a Majorana mass for
the $Y=0$ Dirac triplet and provides the inelastic splitting.

At dimension five the relevant operators preserve $\U(1)_X$,
\begin{equation}
\frac{g_\psi^2}{M_\chi}
\Tr\!\left[
\bar\psi T_L^a\psi (\varepsilon_5 T_R^b + \slashed{\varepsilon}_5 \slashed T_R^b)
\right]
\Tr\!\left[
\mathcal H^\dagger
\frac{\sigma^a}{2}
\mathcal H
\frac{\sigma^b}{2}
\right],
\label{eq:dim5operatorsbreaking}
\end{equation}
where $\slashed T_R^b\equiv\sigma_3T_R^b\sigma_3$ and only the term with $\slashed{\varepsilon}_5 \slashed{T}_R^b$ breaks $\SO(4)$.
Their coefficients can be combined as
$\varepsilon_{\rm spl}=(\varepsilon_5+\slashed{\varepsilon}_5)/2$ and
$\varepsilon_{\rm mix}=(\varepsilon_5-\slashed{\varepsilon}_5)/2$, which control the diagonal mass splitting and neutral-state mixing, respectively, after EWSB.

\begin{figure*}[t]
\centering
\includegraphics[width=0.47\textwidth]{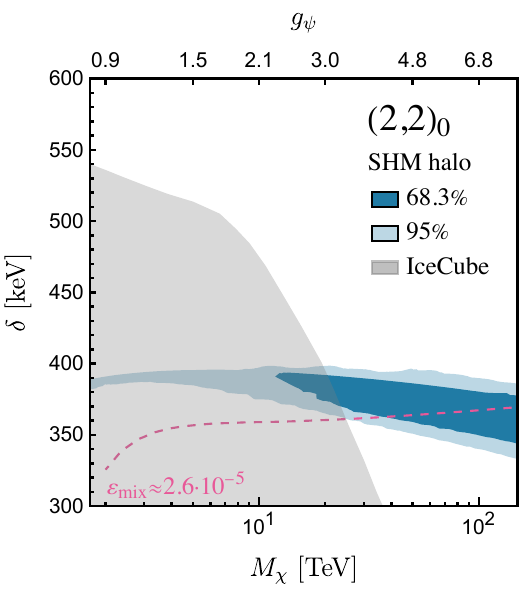}%
\hfill%
\includegraphics[width=0.48\textwidth]{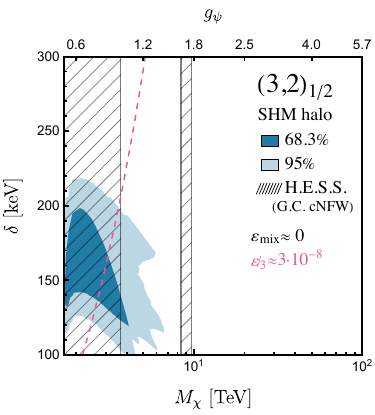}%
\caption{
Thermal interpretations of the LZ230616 recoil event
in the $(M_\chi,\delta)$ plane for the composite Higgsino
$\bid$ (left) and the pure $Y=0$ triplet of the $\trip$
representation (right).
Dark and light blue regions denote the $68.3\%$ and $95\%$
confidence regions obtained from the extended unbinned
likelihood analysis with Monte Carlo calibration,
assuming the SHM.
The gray curve in the left panel shows the IceCube
solar-neutrino constraint, while hatched regions in the
right panel indicate H.E.S.S. gamma-ray line exclusions
for a cNFW Galactic halo profile.
Pink dashed contours correspond to
$\varepsilon_{\rm mix}=2.6\times10^{-5}$ for the Higgsino
and $\slashed{\varepsilon_3}=3\times10^{-8}$ for the
pure triplet. In both panels, the upper x-axis indicates the strong coupling $g_\psi$ corresponding to the DM mass $M_\chi$.
}
\label{fig:results}
\end{figure*}

For the real $\bid$, $\varepsilon_{\rm mix}$ induces maximal mixing between the two neutral components of the Dirac fermion. The resulting Majorana mass eigenstates are split by
\begin{equation}
\delta=\frac{\xi M_\chi}{2}
|\varepsilon_{\rm mix}| \ .
\label{eq:delta22}
\end{equation}
$\varepsilon_{\rm spl}$ instead contributes to the
charged--neutral splitting.

For the $\trip$, the same dimension-five interactions mix the neutral Dirac states belonging to the $Y=0$ and $Y=1$ triplets. Their mixing angle is
\begin{equation}
\tan(2\theta)=-\frac{\xi M_\chi\varepsilon_{\rm mix}}
{\sqrt2\,\Delta_{10}} \ .
\label{eq:theta32}
\end{equation}
The angle $\theta$ therefore directly controls the hypercharge admixture of the light state and, consequently, its coupling to the $Z$ boson. We focus on the branch $\Delta_{10}>0$, for which $|\theta|<\pi/4$ and the light state is predominantly $Y=0$; the opposite ordering gives an EW-dominated interaction similar to the Higgsino case.

These dimension-five interactions preserve $\U(1)_X$ and thus do not split the light Dirac state into Majorana components. The inelastic
threshold is instead generated by the $\U(1)_X$-violating
dimension-three contribution parametrized by
$\slashed{\varepsilon}_3$. Since it acts on the $\mathbf3_0$
component, its projection onto the light state gives
\begin{equation}
\delta\simeq
2|\slashed{\varepsilon}_3|M_\chi\cos^2\theta \ .
\label{eq:delta32}
\end{equation}
Thus, in the $\trip$ benchmark, $\theta$ controls the EW admixture and
the transition strength, while $\slashed{\varepsilon}_3$ independently
sets the inelastic threshold.

The zero-momentum cross sections per neutron for the transition between the two states of the light Majorana pair are
\begin{align}
\sigma_n&=\sigma_n^{\rm EW}
\left(1-\frac{\xi}2\right)^2 & &\text{for}~ \bid \ ,
\label{eq:sigma22}\\
\sigma_n&=\sigma_n^{\rm EW}
\left[2\sin^2\theta+\frac{\xi}4
(1-4\sin^2\theta)\right]^2 & &\text{for}~ \trip \ ,
\label{eq:sigma32}
\end{align}
where $\sigma_n^{\rm EW}=G_F^2\mu_{\chi n}^2/(2\pi) \simeq7.43\times10^{-39}\,\mathrm{cm}^2$ for $M_\chi\gg m_n$. Further details are given in Appendix~\ref{app:spectrum-XS}.
The $\bid$ retains an essentially EW-strength transition, as for a standard Higgsino, whereas in the pure $Y=0$ limit ($\sin\theta\to0$) the $\trip$ cross section is suppressed by $\xi^2/16$. A mixed $\trip$ interpolates between these two regimes. Their distinct DD phenomenology selects markedly different regions of parameter space.

\medskip
\para{The LZ recoil and statistical analysis}
For an endothermic transition, the minimum incoming DM speed required to produce a nuclear recoil of energy $E_R$ is
\begin{equation}
v_{\min}^{A}(E_R)=
\frac{1}{\sqrt{2m_AE_R}}
\left(
\frac{m_AE_R}{\mu_{\chi A}}+\delta
\right),
\label{eq:vmin}
\end{equation}
where $m_A$ and $\mu_{\chi A}$ are the nuclear mass and the DM--nucleus reduced mass, respectively. The minimum occurs at $E_R^\star=\mu_{\chi A}\delta/m_A$, approaching $E_R^\star\simeq\delta$ for $M_\chi\gg m_A$. Increasing the splitting therefore suppresses the accessible halo population while shifting the characteristic recoil energy towards higher values. As a result, the interplay between $\delta$ and the mixing-dependent cross sections in Eq.~\eqref{eq:sigma22}~and Eq.~\eqref{eq:sigma32} determines the distinct recoil spectra of the two benchmarks.

For a xenon isotope $A$, the differential recoil rate per unit detector mass reads
\begin{equation}
\frac{dR_A}{dE_R}
=
\mathcal N_A
\frac{\rho_\chi}{M_\chi}
\frac{m_A\sigma_A}{2\mu_{\chi A}^2}
F_A^2(E_R)\,
\overline\eta\!\left(v_{\min}^{A}(E_R)\right),
\label{eq:isotope}
\end{equation}
where $\mathcal N_A$ accounts for the isotopic abundance and the number of target nuclei per unit detector mass, $F_A(E_R)$ is the isotope-dependent nuclear response extracted from~\cite{Vietze:2014blah}, and $\overline\eta$ is the annually averaged velocity integral computed assuming a Standard Halo Model (SHM) with $\rho_\chi=0.3\,\mathrm{GeV\,cm^{-3}}$, $v_0=238\,\mathrm{km\,s^{-1}}$, and $v_{\rm esc}=544\,\mathrm{km\,s^{-1}}$~\cite{Baxter:2021pqo}. The nuclear cross-section is $\sigma_A=(\mu_{\chi A}/\mu_{\chi n})^2Q_W^2\sigma_n$, with $Q_W=(A-Z)-(1-4s_W^2)Z$. The total spectrum is obtained by summing over the xenon isotopes. We adopt the LZ nuclear-recoil efficiency~\cite{LZ:2026axp}.

\begin{figure*}[t!]
\centering
\includegraphics[width=.53\textwidth]
  {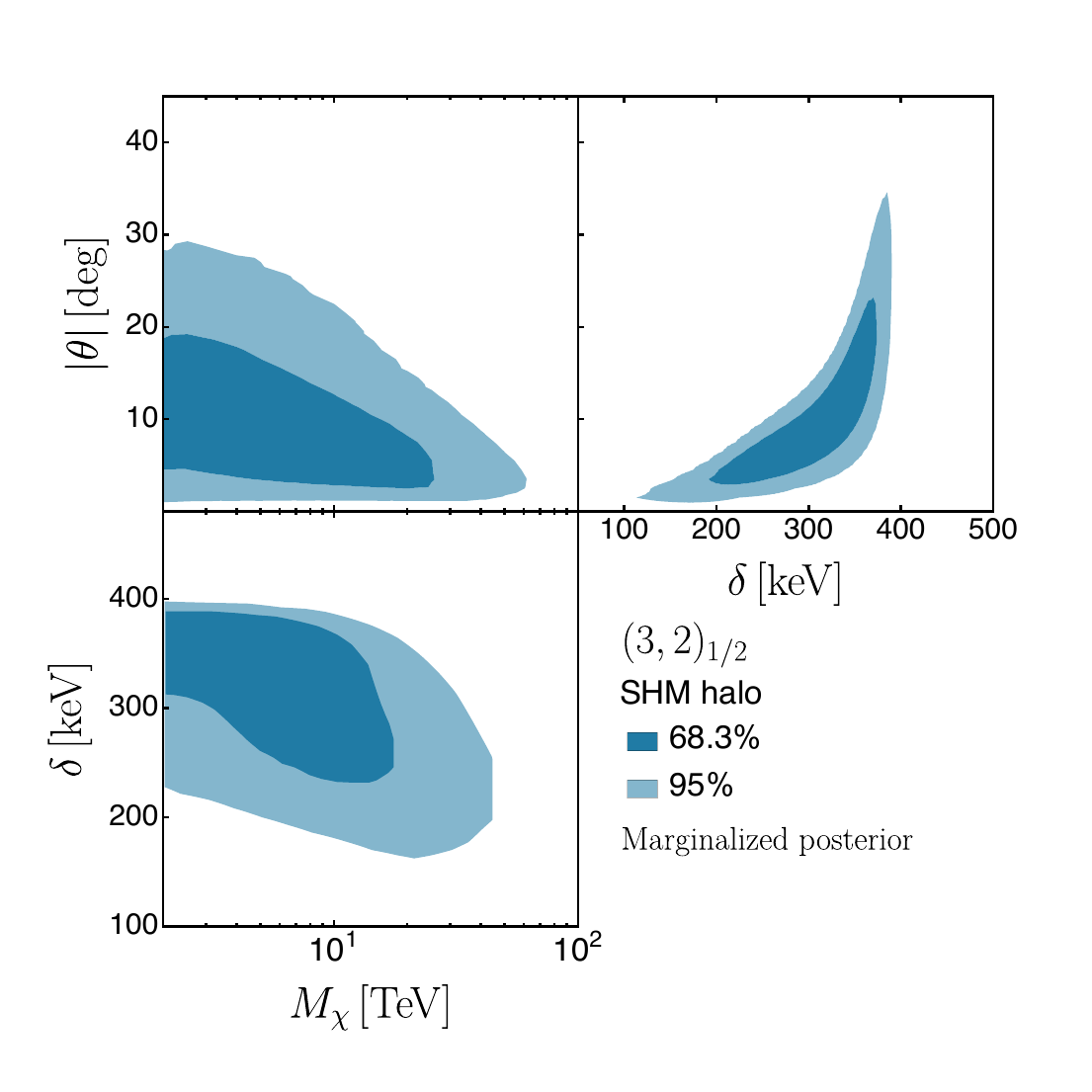}\hfill%
\includegraphics[width=.47\textwidth]{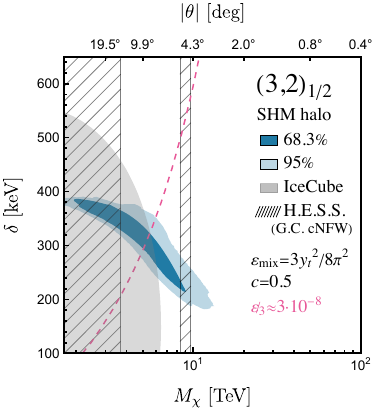}
\caption{
Interpretations of the LZ230616 recoil event for the thermal
mixed $\trip$ scenario, assuming the SHM.
Left: two-dimensional marginalized posteriors in the
$(M_\chi,|\theta|)$ plane (upper left),
the $(\delta,|\theta|)$ plane (upper right),
and the $(M_\chi,\delta)$ plane (lower left).
Dark and light blue denote the $68.3\%$ and $95\%$
highest-posterior-density regions, respectively.
Right: thermal interpretation in the $(M_\chi,\delta)$
plane for the benchmark
$\varepsilon_{\rm mix}=3y_t^2/(8\pi^2)$ and $c=0.5$.
Confidence regions and indirect-detection constraints
follow the conventions of Fig.~\ref{fig:results}.
The upper axis gives the mixing angle $|\theta|$.
The pink dashed contour corresponds to
$\slashed{\varepsilon_3}=3\times10^{-8}$, with the
physical splitting given by
$\delta\simeq2|\slashed{\varepsilon_3}|M_\chi\cos^2\theta$.
}

\label{fig:posterior}
\end{figure*}
As emphasized in Ref.~\cite{Rodd:2026tyn}, the absence of additional high-energy recoil candidates can provide complementary information on inelastic DM scenarios, whose spectra may extend beyond the nominal search window. Motivated by this observation, we consider two disjoint regions: the LZ region of interest (ROI), extending to approximately $270\,\mathrm{keV}$, and a high-energy sideband (HSB) spanning $350\lesssim E_R/\mathrm{keV}\lesssim675$. In the ROI, the predicted spectrum is convolved with a Gaussian energy response and weighted by the LZ nuclear-recoil acceptance. In the HSB, we adopt unit acceptance and neglect energy smearing. Both contributions are normalized to an exposure $\mathcal E=2.84\,\mathrm{tonne\,yr}$. 

Assuming negligible backgrounds, we construct an extended unbinned likelihood for the observed candidate at $E_{\rm obs}=248\,\mathrm{keV}$ in the ROI and zero events in the HSB,
\begin{equation}
\log\mathcal L(\boldsymbol{\varphi})
=-N_{\rm sig}(\boldsymbol{\varphi})
+\log\lambda(E_{\rm obs};\boldsymbol{\varphi}) \ .
\label{eq:lz_likelihood}
\end{equation}
Here $\lambda(E;\boldsymbol{\varphi})$ denotes the differential event yield in the ROI, while $N_{\rm sig}=N_{\rm ROI}+N_{\rm HSB}$ is the total expected yield. For the $\bid$ benchmark $\varphi=\{M_\chi,\delta\}$, while for the mixed $\trip$ scenario $\boldsymbol{\varphi}=\{M_\chi,\delta,\varepsilon_{\rm mix},c\}$. The first term in Eq.~\eqref{eq:lz_likelihood} accounts for the total event yield, including the contribution from the empty HSB, while the second incorporates the recoil-energy information of the observed event. We construct confidence regions using a likelihood-ratio test calibrated with Monte Carlo pseudo-experiments, due to the low number of events. Details regarding the detector-response and the statistical procedure are provided in Appendix~\ref{app:recast}.

The left panel of Fig.~\ref{fig:results} shows the thermal composite Higgsino $\bid$, with dark- and light-blue regions corresponding to $68.3\%$ and $95\%$ confidence levels, respectively. Its EW cross section in Eq.~\eqref{eq:sigma22} requires a large mass splitting, $\delta\gtrsim370\,\mathrm{keV}$, to suppress the scattering rate. Indeed, at such splitting only DM particles in the high-velocity tail of the halo distribution scatter. The observed recoil energy and the absence of events in the HSB jointly constrain the allowed region, while increasing $M_\chi$ further suppresses the event yield through the reduced DM flux. 
The pink dashed contour shows the splitting induced by $\U(1)_X$-preserving dimension-five interactions, with the preferred region probing $\varepsilon_{\rm mix}\sim\mathrm{few}\times10^{-5}$.

The pure $Y=0$ triplet of the $\trip$ representation exhibits a qualitatively different behavior. In the absence of mixing, its inelastic $Z$ coupling arises solely from strong-sector dynamics, yielding a suppressed cross section, see Eq.~\eqref{eq:sigma32}. This allows the LZ event to be accommodated at smaller splittings, $\delta\sim150~\mathrm{keV}$, probing a less extreme portion of the halo velocity distribution, as shown in the right panel of Fig.~\ref{fig:results}. 
Unlike the $\bid$ case, the splitting originates from a dimension-three operator explicitly breaking $\U(1)_X$, with the pink dashed contour corresponding to $\slashed{\varepsilon}_3\sim\mathrm{few}\times10^{-8}$. Such a small coefficient suggests that $\U(1)_X$ is an approximate symmetry of the strong sector, whose weak breaking naturally accounts for the small inelastic mass splitting. In realizations where the bottom Yukawa breaks $\U(1)_X$, as in the bidoublet model of Ref.~\cite{SO4neutralinos_DiLecce}, reproducing the required splitting demands additional suppression.

\smallskip
Beyond the pure $Y=0$ limit, mixing with the $Y=1$ component introduces an additional EW contribution, enhancing the $Z$-mediated inelastic transition. The mixing angle $\theta(M_\chi,\varepsilon_{\rm mix},c)$ is controlled by an interplay between the EW mass difference and the $\SO(5)$-breaking couplings $\varepsilon_3, \varepsilon_{\rm mix}$. It determines the scattering cross section, while $\slashed{\varepsilon}_3$ sets the Majorana splitting $\delta$ through Eq.~\eqref{eq:delta32}.

To assess the sensitivity of the LZ event to the underlying mixing, we also perform a Bayesian analysis of the mixed $\trip$ scenario. The marginalized posteriors in the left panel of Fig.~\ref{fig:posterior} reveal correlations among $M_\chi$, $\delta$, and $|\theta|$, with a broad preference for mixing angles $|\theta|\sim10^\circ$, DM masses up to $M_\chi\lesssim 20~\rm TeV$ and $200\lesssim\delta/{\rm keV}\lesssim380$. The dark- and light-blue regions enclose $68.3\%$ and $95\%$ of the marginalized posterior probability, respectively. Clearly, they have a different statistical meaning with respect to the frequentist confidence regions of Fig.~\ref{fig:results}. Details are given in Appendix~\ref{app:recast}.

\smallskip
Interestingly, a symmetry-breaking interaction generated radiatively by the top sector can induce a large enough mixing angle for TeV scale DM masses and $c\sim{\cal O}(1)$. In the right panel of Fig.~\ref{fig:posterior} we show the confidence regions for the corresponding benchmark $\varepsilon_{\rm mix}=3y_t^2/(8\pi^2)$ and $c=0.5$.
These values should be understood as parametric estimates, with undetermined order-one coefficients. The preferred region lies around $M_\chi\sim6\,\mathrm{TeV}$ and $\delta\sim300\,\mathrm{keV}$, corresponding to $|\theta|\sim10^\circ$. 

\para{Complementary bounds}
The Sun provides a particularly powerful probe of inelastic Majorana pairs with EW-scale scattering cross sections~\cite{Menon:2009qj,Nussinov:2009ft}.
DM captured in the Sun and annihilating into $WW$ can be constrained by neutrino detectors beyond the reach of DD experiments~\cite{Bose:2026ndd,Pospelov:2026ewn,Luengas:2026tlf}. Following Refs.~\cite{Luengas:2026tlf,Pospelov:2026ewn}, the recast of the IceCube bound~\cite{IceCube:2025fcu} gives the exclusions shown in gray in Fig.~\ref{fig:results} (left) and Fig.~\ref{fig:posterior} (right). For the $\bid$ it reaches up to $M_\chi\sim 20~\,\mathrm{TeV}$, not affecting the LZ best-fit region. Capture is inefficient for a pure $Y=0$ triplet, while solar-neutrino searches constrain part of the mixed $\trip$ parameter space, with weaker bounds than for the Higgsino due to the suppressed inelastic cross section. The IceCube recast excludes masses below about $5\,\mathrm{TeV}$. Details are given in Appendix~\ref{app:sc}.

\smallskip
Gamma-ray searches provide powerful indirect probes of EW-like DM, particularly in the multi-TeV regime~\cite{Cirelli:2015bda,Cirelli:2008pk,Rinchiuso:2018ajn,Baumgart:2023quintuplet,Aghaie:2025iyn}. Present-day indirect detection probes $\chi_0\chi_0$ annihilation, in contrast to the thermal average over the full multiplet involved in freeze-out. The leading strong-sector operator in Eq.~\eqref{eq:strong} generates no tree-level $s$-wave contribution for this initial state, so that the present-day signal is governed by EW interactions even when strong dynamics control the relic abundance~\cite{SO4neutralinos_DiLecce}. We compare the predicted $\gamma$-ray signal with the H.E.S.S. Inner Galaxy Survey line-search limits based on $546\,\mathrm{h}$ of observations~\cite{HESS:2026ugs}. These constraints are negligible for the $\bid$ over the mass range surviving solar-neutrino searches, but are relevant for the few-TeV regions compatible with LZ in the pure and mixed $\trip$ scenarios. Since these states are predominantly Wino-like, we approximate the hard-photon spectrum using the Sommerfeld-enhanced real-triplet result for $\chi^0\chi^0\to\gamma+X$ at next-to-leading-logarithmic accuracy~\cite{Baumgart:2018yed,Baumgart:2025dov}. Assuming a cNFW profile with a $1\,\mathrm{kpc}$ core, the resulting exclusions, hatched in the right panels of Figs.~\ref{fig:results} and~\ref{fig:posterior}, rule out $M_\chi\lesssim4\,\mathrm{TeV}$ and a narrow interval around $M_\chi\simeq9\,\mathrm{TeV}$.

\medskip
\para{Conclusions}
We have presented the first thermal inelastic DM interpretation of
LZ230616 embedded in the Minimal Composite Higgs Model, a UV framework
motivated by the EW hierarchy problem. DM arises as
the lightest fermionic composite resonance and is stabilized by an
accidental parity of the strong sector, without requiring an {\it ad-hoc}
discrete symmetry. Composite interactions open additional annihilation
channels, extending the thermal mass range beyond the pure EW prediction,
while the relic abundance relates the DM mass to a single strong coupling
and preserves the predictivity of the framework.

Two representations realize three distinct regimes compatible with the
LZ high-energy sideband and the solar-neutrino and gamma-ray
bounds. The real bidoublet has Higgsino-like EW content and an
unsuppressed inelastic $Z$ coupling, requiring $M_\chi\gtrsim20\,\mathrm{TeV}$ and $\delta\sim370\,\mathrm{keV}$. Remarkably, such masses correspond to $g_\psi\gtrsim 3$, typical of strongly-coupled theories.
The complex $\trip$ realizes two qualitatively different possibilities. For a pure $Y=0$ triplet, the strong dynamics generates the inelastic cross section, leaving only a tiny viable region at few-TeV masses and $\delta\sim150\,\mathrm{keV}$. Mixing with the hypercharged triplet continuously enhances the transition strength; a moderate mixing $|\theta|\sim10^\circ$ yields a viable thermal benchmark near $M_\chi\sim6\,\mathrm{TeV}$ and $\delta\sim300\,\mathrm{keV}$. This mixing angle is remarkably compatible with the size expected from radiative $\SO(5)$ breaking induced by the top quark. The recoil rate and the inelastic threshold probe different symmetry-breaking operators of the composite sector: for the $\trip$, the transition strength probes the $\SO(5)$-breaking mixing between states of different hypercharge, whereas the small Majorana splitting probes the breaking of $\U(1)_X$. 

A confirmation or exclusion of LZ230616 with forthcoming nuclear-recoil data would therefore probe the hierarchy of symmetry-breaking interactions within a predictive thermal DM framework motivated by electroweak naturalness, while remaining consistent with current complementary constraints.

\balance
\bibliographystyle{JHEP}
\bibliography{biblio}

@preamble{"\providecommand{\noopsort}[1]{}" #
   "\providecommand{\singleletter}[1]{#1}%"}

@article{Bottaro:2022one,
	archiveprefix = {arXiv},
	author = {Bottaro, Salvatore and Buttazzo, Dario and Costa, Marco and Franceschini, Roberto and Panci, Paolo and Redigolo, Diego and Vittorio, Ludovico},
	doi = {10.1140/epjc/s10052-022-10918-5},
	eprint = {2205.04486},
	journal = {Eur. Phys. J. C},
	number = {11},
	pages = {992},
	primaryclass = {hep-ph},
	reportnumber = {CERN-TH-2022-080},
	title = {{The last complex WIMPs standing}},
	volume = {82},
	year = {2022}}

@article{LZ:2026axp,
	archiveprefix = {arXiv},
	author = {Akerib, D. S. and others},
	collaboration = {LZ},
	eprint = {2609.02823},
	month = {9},
	primaryclass = {hep-ex},
	title = {{Search for dark matter particle interactions in an extended nuclear recoil energy window with the LUX-ZEPLIN (LZ) experiment}},
	year = {2026}}

@article{Cirelli:2005uq,
	archiveprefix = {arXiv},
	author = {Cirelli, Marco and Fornengo, Nicolao and Strumia, Alessandro},
	doi = {10.1016/j.nuclphysb.2006.07.012},
	eprint = {hep-ph/0512090},
	journal = {Nucl. Phys. B},
	pages = {178--194},
	reportnumber = {DFTT40-2005, IFUP-TH-2005-34},
	title = {{Minimal dark matter}},
	volume = {753},
	year = {2006}}

@article{Aghaie:2025iyn,
	archiveprefix = {arXiv},
	author = {Aghaie, Mohammad and Dondarini, Alessandro and Marino, Giulio and Panci, Paolo},
	doi = {10.1088/1475-7516/2026/07/063},
	eprint = {2507.17607},
	journal = {JCAP},
	pages = {063},
	primaryclass = {hep-ph},
	title = {{Minimal Dark Matter in the sky: updated Indirect Detection probes}},
	volume = {07},
	year = {2026}}

@article{Fan:2026kxx,
	archiveprefix = {arXiv},
	author = {Fan, JiJi and Reece, Matthew},
	eprint = {2609.01504},
	month = {9},
	primaryclass = {hep-ph},
	title = {{Higgsino Above the Sea of Fog}},
	year = {2026}}

@article{Freese:2026sga,
	archiveprefix = {arXiv},
	author = {Freese, Katherine and Theodosopoulos, Dionysios P.},
	eprint = {2609.01583},
	month = {9},
	primaryclass = {hep-ph},
	title = {{Higgsino Dark Matter Interpretation of the LUX-ZEPLIN 248 keV Nuclear-Recoil Event}},
	year = {2026}}

@article{Wu:2026nhi,
	archiveprefix = {arXiv},
	author = {Wu, Lei and Zhang, Yang and Zhu, Bin},
	eprint = {2609.01590},
	month = {9},
	primaryclass = {hep-ph},
	title = {{TeV Higgsino Dark Matter from LZ Nuclear Recoil to Fermi-LAT Gamma Rays}},
	year = {2026}}

@article{Du:2026guj,
	archiveprefix = {arXiv},
	author = {Du, Xiaokang and Wang, Fei},
	eprint = {2609.04163},
	month = {9},
	primaryclass = {hep-ph},
	title = {{TeV Higgsino Interpretation of the LZ High-Recoil Event with Intermediate-Scale Electroweak Gauginos}},
	year = {2026}}

@article{Pospelov:2026ewn,
	archiveprefix = {arXiv},
	author = {Pospelov, Maxim and Ramani, Harikrishnan},
	eprint = {2609.02775},
	month = {9},
	primaryclass = {hep-ph},
	title = {{Strong Constraints on Higgsino Dark Matter from Solar Capture}},
	year = {2026}}

@article{Tucker-Smith:2001myb,
	archiveprefix = {arXiv},
	author = {Tucker-Smith, David and Weiner, Neal},
	doi = {10.1103/PhysRevD.64.043502},
	eprint = {hep-ph/0101138},
	journal = {Phys. Rev. D},
	pages = {043502},
	reportnumber = {UCB-PTH-00-43, LBNL-47234, UW-PT-00-17},
	title = {{Inelastic dark matter}},
	volume = {64},
	year = {2001}}

@article{Kaplan:1991dc,
    author = "Kaplan, David B.",
    title = "{Flavor at SSC energies: A New mechanism for dynamically generated fermion masses}",
    reportNumber = "UCSD-PTH-91-04",
    doi = "10.1016/S0550-3213(05)80021-5",
    journal = "Nucl. Phys. B",
    volume = "365",
    pages = "259--278",
    year = "1991"
}

@article{Vietze:2014blah,
	archiveprefix = {arXiv},
	author = {Vietze, L. and Klos, P. and Men{\'e}ndez, J. and Haxton, W. C. and Schwenk, A.},
	doi = {10.1103/PhysRevD.91.043520},
	eprint = {1412.6091},
	journal = {Phys. Rev. D},
	number = {4},
	pages = {043520},
	primaryclass = {nucl-th},
	title = {{Nuclear structure aspects of spin-independent WIMP scattering off xenon}},
	volume = {91},
	year = {2015}}

@article{IceCube:2025fcu,
	archiveprefix = {arXiv},
	author = {Abbasi, R. and others},
	collaboration = {IceCube},
	eprint = {2507.08457},
	month = {7},
	primaryclass = {hep-ex},
	title = {{Search for High-Energy Neutrinos From the Sun Using Ten Years of IceCube Data}},
	year = {2025}}

@article{Rodd:2026tyn,
	archiveprefix = {arXiv},
	author = {Rodd, Nicholas L. and Safdi, Benjamin R. and Slatyer, Tracy R. and Xu, Weishuang Linda},
	eprint = {2609.04175},
	month = {9},
	primaryclass = {hep-ph},
	title = {{Confronting the Higgsino Interpretation of the LZ Event with the High-Energy Sideband}},
	year = {2026}}

@article{Schoenrich:2009bx,
	author = {Sch{\"o}nrich, Ralph and Binney, James and Dehnen, Walter},
	doi = {10.1111/j.1365-2966.2010.16253.x},
	issn = {1365-2966},
	journal = {Monthly Notices of the Royal Astronomical Society},
	month = Apr,
	number = {4},
	pages = {1829--1833},
	publisher = {Oxford University Press (OUP)},
	title = {Local kinematics and the local standard of rest},
	url = {http://dx.doi.org/10.1111/j.1365-2966.2010.16253.x},
	volume = {403},
	year = {2010}}

@article{Bose:2026ndd,
	archiveprefix = {arXiv},
	author = {Bose, Debajit and others},
	eprint = {2609.07807},
	month = {9},
	primaryclass = {hep-ph},
	title = {{Not so good $\nu$s for Higgsino dark matter as LZ excess: stringent limits from Super-Kamiokande and IceCube}},
	year = {2026}}

@article{Helm:1956zz,
	author = {Helm, Richard H.},
	doi = {10.1103/PhysRev.104.1466},
	journal = {Phys. Rev.},
	pages = {1466--1475},
	title = {{Inelastic and Elastic Scattering of 187-Mev Electrons from Selected Even-Even Nuclei}},
	volume = {104},
	year = {1956}}

@article{Lewin:1995rx,
	author = {Lewin, J. D. and Smith, P. F.},
	doi = {10.1016/S0927-6505(96)00047-3},
	journal = {Astropart. Phys.},
	pages = {87--112},
	title = {{Review of mathematics, numerical factors, and corrections for dark matter experiments based on elastic nuclear recoil}},
	volume = {6},
	year = {1996}}

@article{Meija:2016,
	author = {Meija, Juris and Coplen, Tyler B. and Berglund, Michael and others},
	doi = {10.1515/pac-2015-0503},
	journal = {Pure Appl. Chem.},
	number = {3},
	pages = {293--306},
	title = {{Isotopic compositions of the elements 2013 (IUPAC Technical Report)}},
	volume = {88},
	year = {2016}}

@article{HESS:2026ugs,
	archiveprefix = {arXiv},
	author = {{H.E.S.S. Collaboration}},
	eprint = {2608.07234},
	month = {8},
	primaryclass = {astro-ph.HE},
	title = {{Search for gamma-ray spectral lines from dark matter annihilation with the H.E.S.S. Inner Galaxy Survey}},
	year = {2026}}

@unpublished{SO4neutralinos_DiLecce,
	author = {Buttazzo, Dario and Di Lecce, Andrea and Franceschini, Roberto and Panci, Paolo},
	note = {In preparation},
	title = {The (Partially) Composite Neutralinos},
	year = {2026}}

@article{FeldmanCousins,
	archiveprefix = {arXiv},
	author = {Feldman, Gary J. and Cousins, Robert D.},
	doi = {10.1103/PhysRevD.57.3873},
	eprint = {physics/9711021},
	journal = {Phys. Rev. D},
	pages = {3873--3889},
	reportnumber = {HUTP-97-A096},
	title = {{A Unified approach to the classical statistical analysis of small signals}},
	volume = {57},
	year = {1998}}

@article{Neyman1937,
	author = {Neyman, Jerzy},
	doi = {10.1098/rsta.1937.0005},
	journal = {Phil. Trans. Roy. Soc. Lond. A},
	pages = {333--380},
	title = {{Outline of a Theory of Statistical Estimation Based on the Classical Theory of Probability}},
	volume = {236},
	year = {1937}}

@article{Wilks1938,
	author = {Wilks, Samuel S.},
	doi = {10.1214/aoms/1177732360},
	journal = {Annals Math. Statist.},
	number = {1},
	pages = {60--62},
	title = {{The Large-Sample Distribution of the Likelihood Ratio for Testing Composite Hypotheses}},
	volume = {9},
	year = {1938}}

@article{Cowan:2010js,
	archiveprefix = {arXiv},
	author = {Cowan, Glen and Cranmer, Kyle and Gross, Eilam and Vitells, Ofer},
	doi = {10.1140/epjc/s10052-011-1554-0},
	eprint = {1007.1727},
	journal = {Eur. Phys. J. C},
	note = {[Erratum: Eur. Phys. J. C 73 (2013) 2501]},
	pages = {1554},
	primaryclass = {physics.data-an},
	title = {{Asymptotic formulae for likelihood-based tests of new physics}},
	volume = {71},
	year = {2011}}

@article{mchm,
	archiveprefix = {arXiv},
	author = {Agashe, Kaustubh and Contino, Roberto and Pomarol, Alex},
	doi = {10.1016/j.nuclphysb.2005.04.035},
	eprint = {hep-ph/0412089},
	journal = {Nucl. Phys. B},
	pages = {165--187},
	title = {{The Minimal Composite Higgs Model}},
	volume = {719},
	year = {2005}}

@article{Yamada_2010,
	author = {Yamada, Youichi},
	doi = {10.1016/j.physletb.2009.11.044},
	issn = {0370-2693},
	journal = {Physics Letters B},
	month = Jan,
	number = {4-5},
	pages = {435--440},
	publisher = {Elsevier BV},
	title = {Electroweak two-loop contribution to the mass splitting within a new heavy fermion multiplet},
	url = {http://dx.doi.org/10.1016/j.physletb.2009.11.044},
	volume = {682},
	year = {2010}}

@article{Cirelli:2007xd,
	archiveprefix = {arXiv},
	author = {Cirelli, Marco and Strumia, Alessandro and Tamburini, Matteo},
	doi = {10.1016/j.nuclphysb.2007.07.023},
	eprint = {0706.4071},
	journal = {Nucl. Phys. B},
	pages = {152--175},
	primaryclass = {hep-ph},
	reportnumber = {IFUP-TH-2007-15, SACLAY-T07-064},
	title = {{Cosmology and Astrophysics of Minimal Dark Matter}},
	volume = {787},
	year = {2007}}

@article{Cirelli:2008pk,
	archiveprefix = {arXiv},
	author = {Cirelli, Marco and Franceschini, Roberto and Strumia, Alessandro},
	doi = {10.1016/j.nuclphysb.2008.03.013},
	eprint = {0802.3378},
	journal = {Nucl. Phys. B},
	pages = {204--220},
	primaryclass = {hep-ph},
	reportnumber = {IFUP-TH-2008-05, SACLAY-T08-034},
	title = {{Minimal Dark Matter predictions for galactic positrons, anti-protons and photons}},
	volume = {800},
	year = {2008}}

@article{DelNobile:2015bqo,
	archiveprefix = {arXiv},
	author = {Del Nobile, Eugenio and Nardecchia, Marco and Panci, Paolo},
	doi = {10.1088/1475-7516/2016/04/048},
	eprint = {1512.05353},
	journal = {JCAP},
	pages = {048},
	primaryclass = {hep-ph},
	reportnumber = {DFPD-2015-TH-29},
	title = {{Millicharge or Decay: A Critical Take on Minimal Dark Matter}},
	volume = {04},
	year = {2016}}

@article{Bottaro:2021snn,
	archiveprefix = {arXiv},
	author = {Bottaro, Salvatore and Buttazzo, Dario and Costa, Marco and Franceschini, Roberto and Panci, Paolo and Redigolo, Diego and Vittorio, Ludovico},
	doi = {10.1140/epjc/s10052-021-09917-9},
	eprint = {2107.09688},
	journal = {Eur. Phys. J. C},
	number = {1},
	pages = {31},
	primaryclass = {hep-ph},
	title = {{Closing the Window on WIMP Dark Matter}},
	volume = {82},
	year = {2022}}

@article{Panci:2024ewmultiplets,
	archiveprefix = {arXiv},
	author = {Panci, Paolo},
	eprint = {2405.05087},
	journal = {PoS},
	pages = {033},
	primaryclass = {hep-ph},
	title = {{Electroweak Multiplets as Dark Matter Candidates: A Brief Review}},
	volume = {CORFU2023},
	year = {2024}}

@article{Cirelli:2024darkmatter,
	archiveprefix = {arXiv},
	author = {Cirelli, Marco and Strumia, Alessandro and Zupan, Jure},
	eprint = {2406.01705},
	month = {6},
	primaryclass = {hep-ph},
	title = {{Dark Matter}},
	year = {2024}}

@article{Baumgart:2017nsr,
	archiveprefix = {arXiv},
	author = {Baumgart, Matthew and Cohen, Timothy and Moult, Ian and Rodd, Nicholas L. and Slatyer, Tracy R. and Solon, Mikhail P. and Stewart, Iain W. and Vaidya, Varun},
	doi = {10.1007/JHEP03(2018)117},
	eprint = {1712.07656},
	journal = {JHEP},
	pages = {117},
	primaryclass = {hep-ph},
	reportnumber = {CALT-TH-2017-066, LA-UR-17-31169, MIT-CTP-4959},
	title = {{Resummed Photon Spectra for WIMP Annihilation}},
	volume = {03},
	year = {2018}}

@article{Rinchiuso:2018ajn,
	archiveprefix = {arXiv},
	author = {Rinchiuso, Lucia and Rodd, Nicholas L. and Moult, Ian and Moulin, Emmanuel and Baumgart, Matthew and Cohen, Timothy and Slatyer, Tracy R. and Stewart, Iain W. and Vaidya, Varun},
	doi = {10.1103/PhysRevD.98.123014},
	eprint = {1808.04388},
	journal = {Phys. Rev. D},
	number = {12},
	pages = {123014},
	primaryclass = {astro-ph.HE},
	reportnumber = {IRFU-18-18, MIT-CTP-5041, LA-UR-18-28196},
	title = {{Hunting for Heavy Winos in the Galactic Center}},
	volume = {98},
	year = {2018}}

@article{Baumgart:2023quintuplet,
	archiveprefix = {arXiv},
	author = {Baumgart, Matthew and Rodd, Nicholas L. and Slatyer, Tracy R. and Vaidya, Varun},
	doi = {10.1007/JHEP01(2024)158},
	eprint = {2309.11562},
	journal = {JHEP},
	pages = {158},
	primaryclass = {hep-ph},
	title = {{The Quintuplet Annihilation Spectrum}},
	volume = {01},
	year = {2024}}

@article{Thomas:1998wy,
	archiveprefix = {arXiv},
	author = {Thomas, Scott and Wells, James D.},
	doi = {10.1103/PhysRevLett.81.34},
	eprint = {hep-ph/9804359},
	journal = {Phys. Rev. Lett.},
	pages = {34--37},
	reportnumber = {SLAC-PUB-7799, SU-ITP-98-22},
	title = {{Phenomenology of Massive Vectorlike Doublet Leptons}},
	volume = {81},
	year = {1998}}

@article{Nagata:2014aoa,
	archiveprefix = {arXiv},
	author = {Nagata, Natsumi and Shirai, Satoshi},
	doi = {10.1007/JHEP01(2015)029},
	eprint = {1410.4549},
	journal = {JHEP},
	pages = {029},
	primaryclass = {hep-ph},
	reportnumber = {DESY-14-180, FTPI-MINN-14/37, IPMU14-0320},
	title = {{Higgsino Dark Matter in High-Scale Supersymmetry}},
	volume = {01},
	year = {2015}}

@article{Langhoff:2026ujr,
	archiveprefix = {arXiv},
	author = {Langhoff, Kevin},
	eprint = {2609.09385},
	month = {9},
	primaryclass = {hep-ph},
	title = {{Heavy Higgsino Interpretation of the LZ Event}},
	year = {2026}}

@article{Coleman:1969,
	author = {Coleman, Sidney and Wess, Julius and Zumino, Bruno},
	doi = {10.1103/PhysRev.177.2239},
	journal = {Phys. Rev.},
	pages = {2239--2247},
	title = {{Structure of Phenomenological Lagrangians. I}},
	volume = {177},
	year = {1969}}

@article{Callan:1969,
	author = {Callan, Curtis G. and Coleman, Sidney and Wess, Julius and Zumino, Bruno},
	doi = {10.1103/PhysRev.177.2247},
	journal = {Phys. Rev.},
	pages = {2247--2250},
	title = {{Structure of Phenomenological Lagrangians. II}},
	volume = {177},
	year = {1969}}

@book{Panico:2016,
	archiveprefix = {arXiv},
	author = {Panico, Giuliano and Wulzer, Andrea},
	doi = {10.1007/978-3-319-22617-0},
	eprint = {1506.01961},
	primaryclass = {hep-ph},
	publisher = {Springer},
	series = {Lecture Notes in Physics},
	title = {{The Composite Nambu-Goldstone Higgs}},
	volume = {913},
	year = {2016}}

@article{Baxter:2021pqo,
    author = "Baxter, D. and others",
    title = "{Recommended conventions for reporting results from direct dark matter searches}",
    eprint = "2105.00599",
    archivePrefix = "arXiv",
    primaryClass = "hep-ex",
    doi = "10.1140/epjc/s10052-021-09655-y",
    journal = "Eur. Phys. J. C",
    volume = "81",
    number = "10",
    pages = "907",
    year = "2021"
}

@article{Bloch:2024suj,
    author = "Bloch, Itay M. and Bottaro, Salvatore and Redigolo, Diego and Vittorio, Ludovico",
    title = "{Looking for WIMPs through the neutrino fogs}",
    eprint = "2410.02723",
    archivePrefix = "arXiv",
    primaryClass = "hep-ph",
    doi = "10.1007/JHEP08(2025)216",
    journal = "JHEP",
    volume = "08",
    pages = "216",
    year = "2025"
}

@article{Luengas:2026tlf,
    author = "Luengas, Juan P. and Pospelov, Maxim and Ramani, Harikrishnan",
    title = "{Constraints on Complex Electroweak WIMPs from Solar Capture}",
    eprint = "2609.38316",
    archivePrefix = "arXiv",
    primaryClass = "hep-ph",
    month = "9",
    year = "2026"
}

@article{Baumgart:2025dov,
    author = "Baumgart, Matthew and Bottaro, Salvatore and Redigolo, Diego and Rodd, Nicholas L. and Slatyer, Tracy R.",
    title = "{Testing real WIMPs with CTAO}",
    eprint = "2507.15937",
    archivePrefix = "arXiv",
    primaryClass = "hep-ph",
    reportNumber = "MIT-CTP 5893",
    doi = "10.1007/JHEP02(2026)213",
    journal = "JHEP",
    volume = "02",
    pages = "213",
    year = "2026"
}

@article{Baumgart:2018yed,
    author = "Baumgart, Matthew and Cohen, Timothy and Moulin, Emmanuel and Moult, Ian and Rinchiuso, Lucia and Rodd, Nicholas L. and Slatyer, Tracy R. and Stewart, Iain W. and Vaidya, Varun",
    title = "{Precision Photon Spectra for Wino Annihilation}",
    eprint = "1808.08956",
    archivePrefix = "arXiv",
    primaryClass = "hep-ph",
    reportNumber = "LA-UR-18-25972, MIT-CTP 5025",
    doi = "10.1007/JHEP01(2019)036",
    journal = "JHEP",
    volume = "01",
    pages = "036",
    year = "2019"
}

@article{Bisal:2026lz,
  author = {Subhadip Bisal and Junjie Cao and Fei Li},
  title = {{Higgsino Dark Matter Interpretation of the LZ High-Recoil Event in the GNMSSM with TeV-Scale Gauginos}},
  eprint = {2609.07811},
  archivePrefix = {arXiv},
  primaryClass = {hep-ph},
  year = {2026}
}

@article{Chatterjee:2026lz,
  author = {Arindam Chatterjee and Debottam Das and Syed Adil Pasha and Alexander Pukhov and Rahul Puri},
  title = {{Radiative Corrections to the Direct Detection of Inelastic Scattering of Higgsino-like Neutralino Dark Matter}},
  eprint = {2609.09830},
  archivePrefix = {arXiv},
  primaryClass = {hep-ph},
  year = {2026}
}

@article{Frolovsky:2026lz,
  author = {Daniel Frolovsky and Sergei V. Ketov},
  title = {{Higgsino dark matter in the Starobinsky supergravity with the MSSM in light of the LUX-ZEPLIN event}},
  eprint = {2609.11241},
  archivePrefix = {arXiv},
  primaryClass = {hep-ph},
  year = {2026}
}

@article{Khan:2026susy331,
  author = {Imtiaz Khan and Ali Muhammad and G. Mustafa and Farruh Atamurotov and Ahmadjon Abdujabbarov and Mussawir Khan},
  title = {{LZ-Motivated Pseudo-Dirac Higgsinos in the Supersymmetric 331 Model from the Supersymmetric {SU(6)} GUT Model}},
  eprint = {2609.23691},
  archivePrefix = {arXiv},
  primaryClass = {hep-ph},
  year = {2026}
}

@article{Pallis:2026lz,
  author = {C. Pallis},
  title = {{Heavy Higgsino Dark Matter With Low Reheating in View of the LZ High-Recoil Event}},
  eprint = {2609.33869},
  archivePrefix = {arXiv},
  primaryClass = {hep-ph},
  year = {2026}
}

@article{Khan:2026gmsugra,
  author = {Imtiaz Khan and Waqas Ahmed and Tianjun Li and Shabbar Raza and Ali Muhammad},
  title = {{Thermal Higgsino Dark Matter in GmSUGRA for the LZ Event and Solar-Neutrino Constraint}},
  eprint = {2610.00226},
  archivePrefix = {arXiv},
  primaryClass = {hep-ph},
  year = {2026}
}

@article{Hor:2026lz,
  author = {Shihwen Hor and Natsumi Nagata and Tsutomu T. Yanagida},
  title = {{From the LZ Event to Grand Unification: Heavy Higgsinos and Proton Decay}},
  eprint = {2610.01613},
  archivePrefix = {arXiv},
  primaryClass = {hep-ph},
  year = {2026}
}

@article{Freese:2026icecube,
  author = {Katherine Freese and Dionysios P. Theodosopoulos},
  title = {{Higgsino dark matter compatible with the LUX-ZEPLIN high-energy nuclear-recoil event and IceCube constraints}},
  eprint = {2610.03692},
  archivePrefix = {arXiv},
  primaryClass = {hep-ph},
  year = {2026}
}

@article{Jeong:2026lz,
  author = {Kwang Sik Jeong and Fuminobu Takahashi},
  title = {{Environment-Dependent Annihilation of Axion-Coupled Higgsino Dark Matter}},
  eprint = {2610.05893},
  archivePrefix = {arXiv},
  primaryClass = {hep-ph},
  year = {2026}
}

@article{Wang:2026lz,
  author = {Xintian Wang and Jin Min Yang and Yuanfang Yue and Yang Zhang},
  title = {{Probing LZ-Motivated TeV Higgsino Dark Matter through Heavier Sparticles at LHC}},
  eprint = {2610.16156},
  archivePrefix = {arXiv},
  primaryClass = {hep-ph},
  year = {2026}
}

@article{Kaplan:1983fs,
    author = "Kaplan, David B. and Georgi, Howard",
    title = "{SU(2) x U(1) Breaking by Vacuum Misalignment}",
    reportNumber = "HUTP-83/A069",
    doi = "10.1016/0370-2693(84)91177-8",
    journal = "Phys. Lett. B",
    volume = "136",
    pages = "183--186",
    year = "1984"
}

@article{Kaplan:1983sm,
    author = "Kaplan, David B. and Georgi, Howard and Dimopoulos, Savas",
    title = "{Composite Higgs Scalars}",
    reportNumber = "HUTP-83/A079",
    doi = "10.1016/0370-2693(84)91178-X",
    journal = "Phys. Lett. B",
    volume = "136",
    pages = "187--190",
    year = "1984"
}

@article{Georgi:1984ef,
    author = "Georgi, Howard and Kaplan, David B. and Galison, Peter",
    title = "{Calculation of the Composite Higgs Mass}",
    reportNumber = "HUTP-84/A018",
    doi = "10.1016/0370-2693(84)90823-2",
    journal = "Phys. Lett. B",
    volume = "143",
    pages = "152--154",
    year = "1984"
}

@article{Georgi:1984af,
    author = "Georgi, Howard and Kaplan, David B.",
    title = "{Composite Higgs and Custodial SU(2)}",
    reportNumber = "HUTP-84/A034",
    doi = "10.1016/0370-2693(84)90341-1",
    journal = "Phys. Lett. B",
    volume = "145",
    pages = "216--220",
    year = "1984"
}

@article{Dugan:1984hq,
    author = "Dugan, Michael J. and Georgi, Howard and Kaplan, David B.",
    title = "{Anatomy of a Composite Higgs Model}",
    reportNumber = "HUTP-84/A073",
    doi = "10.1016/0550-3213(85)90221-4",
    journal = "Nucl. Phys. B",
    volume = "254",
    pages = "299--326",
    year = "1985"
}

@article{Cirelli:2015bda,
    author = "Cirelli, Marco and Hambye, Thomas and Panci, Paolo and Sala, Filippo and Taoso, Marco",
    title = "{Gamma ray tests of Minimal Dark Matter}",
    eprint = "1507.05519",
    archivePrefix = "arXiv",
    primaryClass = "hep-ph",
    reportNumber = "SACLAY-T15-130, ULB-PHYS-TH-15-16",
    doi = "10.1088/1475-7516/2015/10/026",
    journal = "JCAP",
    volume = "10",
    pages = "026",
    year = "2015"
}

@article{DeSimone:2012fs,
    author = "De Simone, Andrea and Matsedonskyi, Oleksii
              and Rattazzi, Riccardo and Wulzer, Andrea",
    title = "{A First Top Partner Hunter's Guide}",
    eprint = "1211.5663",
    archivePrefix = "arXiv",
    primaryClass = "hep-ph",
    doi = "10.1007/JHEP04(2013)004",
    journal = "JHEP",
    volume = "04",
    pages = "004",
    year = "2013"
}

@article{Matsedonskyi:2013nsa,
    author = "Matsedonskyi, Oleksii and Panico, Giuliano
              and Wulzer, Andrea",
    title = "{Light Top Partners and Precision Physics}",
    eprint = "1306.4655",
    archivePrefix = "arXiv",
    primaryClass = "hep-ph",
    doi = "10.1007/JHEP10(2013)160",
    journal = "JHEP",
    volume = "10",
    pages = "160",
    year = "2013"
}

@article{Nussinov:2009ft,
    author = "Nussinov, Shmuel and Wang, Lian-Tao and Yavin, Itay",
    title = "{Capture of Inelastic Dark Matter in the Sun}",
    eprint = "0905.1333",
    archivePrefix = "arXiv",
    primaryClass = "hep-ph",
    doi = "10.1088/1475-7516/2009/08/037",
    journal = "JCAP",
    volume = "08",
    pages = "037",
    year = "2009"
}

@article{Menon:2009qj,
    author = "Menon, Arjun and Morris, Rob and Pierce, Aaron and Weiner, Neal",
    title = "{Capture and Indirect Detection of Inelastic Dark Matter}",
    eprint = "0905.1847",
    archivePrefix = "arXiv",
    primaryClass = "hep-ph",
    reportNumber = "MCTP-09-14",
    doi = "10.1103/PhysRevD.82.015011",
    journal = "Phys. Rev. D",
    volume = "82",
    pages = "015011",
    year = "2010"
}

@article{Cheng:1998hc,
    author = "Cheng, Hsin-Chia and Dobrescu, Bogdan A. and Matchev, Konstantin T.",
    title = "{Generic and chiral extensions of the supersymmetric standard model}",
    eprint = "hep-ph/9811316",
    archivePrefix = "arXiv",
    reportNumber = "FERMILAB-PUB-98-357-T",
    doi = "10.1016/S0550-3213(99)00012-7",
    journal = "Nucl. Phys. B",
    volume = "543",
    pages = "47--72",
    year = "1999"
}

@article{Feng:1999fu,
    author = "Feng, Jonathan L. and Moroi, Takeo and Randall, Lisa and Strassler, Matthew and Su, Shu-fang",
    title = "{Discovering supersymmetry at the Tevatron in wino LSP scenarios}",
    eprint = "hep-ph/9904250",
    archivePrefix = "arXiv",
    reportNumber = "IASSNS-HEP-99-19, MIT-CTP-2830, PUPT-1857",
    doi = "10.1103/PhysRevLett.83.1731",
    journal = "Phys. Rev. Lett.",
    volume = "83",
    pages = "1731--1734",
    year = "1999"
}

@article{Gherghetta:1999sw,
    author = "Gherghetta, Tony and Giudice, Gian F. and Wells, James D.",
    title = "{Phenomenological consequences of supersymmetry with anomaly induced masses}",
    eprint = "hep-ph/9904378",
    archivePrefix = "arXiv",
    reportNumber = "CERN-TH-99-104",
    doi = "10.1016/S0550-3213(99)00429-0",
    journal = "Nucl. Phys. B",
    volume = "559",
    pages = "27--47",
    year = "1999"
}

@article{Ibe:2012sx,
    author = "Ibe, Masahiro and Matsumoto, Shigeki and Sato, Ryosuke",
    title = "{Mass Splitting between Charged and Neutral Winos at Two-Loop Level}",
    eprint = "1212.5989",
    archivePrefix = "arXiv",
    primaryClass = "hep-ph",
    reportNumber = "ICRR-REPORT-641-2012-30, IPMU-12-0231, UT-12-44",
    doi = "10.1016/j.physletb.2013.03.015",
    journal = "Phys. Lett. B",
    volume = "721",
    pages = "252--260",
    year = "2013"
}

@article{McKay:2017xlc,
    author = "McKay, James and Scott, Pat",
    title = "{Two-loop mass splittings in electroweak multiplets: winos and minimal dark matter}",
    eprint = "1712.00968",
    archivePrefix = "arXiv",
    primaryClass = "hep-ph",
    doi = "10.1103/PhysRevD.97.055049",
    journal = "Phys. Rev. D",
    volume = "97",
    number = "5",
    pages = "055049",
    year = "2018"
}

\clearpage
\nobalance
\appendix
\setcounter{equation}{0}\renewcommand{\theequation}{A\arabic{equation}}
\renewcommand{\theHequation}{appendixA.\arabic{equation}}
\setcounter{figure}{0}\renewcommand{\thefigure}{A\arabic{figure}}
\renewcommand{\theHfigure}{appendix.\arabic{figure}}
\twocolumn[{
\begin{center}
{\large\bfseries Appendices}\par\vspace{3pt}
\end{center}\vspace{7pt}
}]

\section{Gauge interactions and composite currents}\label{app:currents}
\setcounter{equation}{0}\renewcommand{\theequation}{A\arabic{equation}}
\renewcommand{\theHequation}{appendixA.\arabic{equation}}

The composite fermionic resonances interact with the SM gauge bosons through their EW charges and with the Higgs sector through additional interactions dictated by the strong dynamics. 

We describe a resonance $\psi\sim(\mathbf n,\mathbf m)_X$ as a matrix transforming under the unbroken $\SU(2)_L\times\SU(2)_R\times\U(1)_X$ subgroup as $\psi\to U_L\psi U_R^\dagger$ and $\psi\to e^{iX\alpha_X}\psi$, where $U_L$ and $U_R$ act in the $\mathbf n$ and $\mathbf m$ representations, respectively, and $X$ is the $\U(1)_X$ charge. The generators $T_L^a$ and $T_R^a$ act on the left and right indices of $\psi$, respectively. 

\medskip
The gauge interactions are encoded in the EW covariant derivative, while the additional Higgs-dependent interactions are discussed below. The covariant kinetic term is
\begin{equation}
\mathcal L_{\rm kin}
=
\zeta\,\Tr\!\left[
\bar\psi(i\slashed{D}-M_\chi)\psi
\right] \ ,
\label{eq:app-kinetic}
\end{equation}
where $M_\chi$ is the common resonance mass before symmetry-breaking corrections. The normalization factor is $\zeta=1~(1/2)$ for a complex (real) multiplet; the latter avoids double counting the degrees of freedom related by the reality condition. Here, the EW covariant derivative is $D_\mu\psi
=
\partial_\mu\psi
-ig_2 W_\mu^aT_L^a\psi
+ig_1 B_\mu\psi T_R^3
-ig_1 XB_\mu\psi$, where $W_\mu^a$ and $B_\mu$ are the $\SU(2)_L$ and $\U(1)_Y$ gauge fields, with couplings $g_2$ and $g_1$, respectively. Repeated adjoint indices $a=1,2,3$ are summed over.

Since $\SU(2)_R$ acts on $\psi$ from the right through $U_R^\dagger$, the eigenvalue of the right generator acting on a component of $\psi$ has the opposite sign to the eigenvalue of the corresponding column matrix $T_R^3$.  The EW charges are
\begin{equation}
Y=T_R^3+X \ ,
\qquad
Q=T_L^3+T_R^3+X \ .
\label{eq:app-charges}
\end{equation}
For a neutral component, $Q=0$ implies $T_L^3=-Y$. Its ordinary coupling to the $Z$ boson is therefore
\begin{equation}
\mathcal L_{\rm gauge}
=
-\frac{g_2Y}{c_W}
Z_\mu\bar\psi_Y^0\gamma^\mu\psi_Y^0,
\label{eq:app-neutral-gauge}
\end{equation}
where $\psi_Y^0$ denotes a neutral Dirac component of hypercharge $Y$,
$c_W\equiv\cos\theta_W$, $s_W\equiv\sin\theta_W$ and $\tan\theta_W=g_1/g_2$. 

\medskip
The interactions with the Higgs  bosons are conveniently described using the Callan--Coleman--Wess--Zumino (CCWZ) formalism~\cite{Coleman:1969,Callan:1969}. Following the prescription in~\cite{Panico:2016}, the leading Higgs-dependent strong interaction follows from the
dressed-resonance kinetic term in the CCWZ construction and takes the
form
\begin{equation}
\Lag_{\rm strong}
=
-\frac{\xi}{v^2}
\left(
J_{\psi,L}^{\mu,a}J_{H,L\,\mu}^{a}
+
J_{\psi,R}^{\mu,a}J_{H,R\,\mu}^{a}
\right) \ .
\label{eq:strongapp}
\end{equation}
where $\xi\equiv v^2/f^2=g_\psi^2v^2/M_\chi^2$. The fermionic currents are defined as
\begin{equation}
\begin{split}
&J_{\psi,L}^{\mu,a}
=
\zeta\,\Tr\!\left[
\bar\psi\gamma^\mu T_L^a\psi
\right] \ , \\
&J_{\psi,R}^{\mu,a}
=
-\zeta\,\Tr\!\left[
\bar\psi\gamma^\mu\psi T_R^a
\right] \ .
\label{eq:app-fermion-currents}
\end{split}
\end{equation}
Here $\Tr$ denotes the trace over the custodial-representation indices.
\smallskip
The Higgs currents are
\begin{equation}
\begin{split}
&J_{H,L}^{\mu,a} =
\frac{i}{2}
\Tr\!\left[
\mathcal H^\dagger\sigma^aD^\mu\mathcal H
\right] \ , \\
&
J_{H,R}^{\mu,a} =
-\frac{i}{2}
\Tr\!\left[
\mathcal H^\dagger D^\mu\mathcal H\,\sigma^a
\right]\ ,
\label{eq:app-Higgs-currents}
\end{split}
\end{equation}
where $\mathcal H
=
{1/\sqrt2}\,
\left[
(v+h)\mathbf1_2+i\varphi^a\sigma^a
\right]$ and $\sigma^a$ are Pauli matrices. Here  $h$ is the physical Higgs excitation, and $\varphi^a$ are the three EW
Goldstone fields. The gauge covariant derivative of the Higgs bidoublet is
$D_\mu\mathcal H
=
\partial_\mu\mathcal H
-ig_2W_\mu^at^a\mathcal H
+ig_1B_\mu\mathcal Ht^3$  with $t^a=\sigma^a/2$. In unitary gauge, $\varphi^a=0$, and the neutral
Higgs currents satisfy
\begin{equation}
J_{H,L}^{\mu,3}
=
-J_{H,R}^{\mu,3}
=
\frac{g_2(v+h)^2}{4c_W}Z^\mu \ .
\label{eq:app-Higgs-current-vev}
\end{equation}
Evaluating Eq.~\eqref{eq:strongapp} on the EW vacuum, the induced neutral
interaction is proportional to $J_{\psi,L}^{\mu,3}-J_{\psi,R}^{\mu,3}$. For a neutral Dirac component, $J_{\psi,L}^{\mu,3}-J_{\psi,R}^{\mu,3}
=-(2Y-X)\bar\psi_Y^0\gamma^\mu\psi_Y^0$. The resulting strong-sector contribution is therefore
\begin{equation}
\Lag_{\rm strong}
=
\frac{g_2}{4c_W} \xi \,
(2Y-X)\,
Z_\mu\bar\psi_Y^0\gamma^\mu\psi_Y^0.
\label{eq:app-neutral-current-identities}
\end{equation}
For the real bidoublet $\bid$, with $X=0$ and $Y=\pm1/2$, the neutral current is the usual Higgsino-like EW current, corrected only at relative order $\xi$ by the composite interaction. The situation is qualitatively different for the complex $\trip$. Once the conjugate representation is included, its neutral sector contains a hypercharge-zero Dirac state built from the two $\mathbf3_0$ multiplets and a hypercharged Dirac state built from $\mathbf3_1\oplus\mathbf3_{-1}$. The latter retains an unsuppressed EW coupling to the $Z$ boson, whereas the former has no renormalizable $Z$ current and couples to it only through the composite interaction, with a coupling suppressed by $\xi$. The physical neutral states therefore interpolate, through their mixing, between an EW-strength current and a genuinely compositeness-induced one.

\section{EW corrections and mass operators}\label{app:breaking-op}
\setcounter{equation}{0}\renewcommand{\theequation}{B\arabic{equation}}
\renewcommand{\theHequation}{appendixB.\arabic{equation}}

Mass splittings and mixing among the components of the composite
resonances receive contributions from EW radiative corrections and
from local operators encoding explicit breaking of the
$\SO(5)\times\U(1)_X$ global symmetry of the strong sector.

\medskip
\para{Electroweak radiative corrections}
EW corrections split the masses of the individual resonance components away from their common tree-level mass $M_\chi$, thereby lifting the degeneracy within the multiplet. For a component of the multiplet with isospin $I=(n-1)/2$, hypercharge $Y$, and electric charge $Q$, the one-loop mass shift relative to $M_\chi$ is~\cite{Yamada_2010}
\begin{equation}
\begin{split}
& \delta M_{\rm EW}^{(Q,Y)} \equiv M_Y^Q-M_\chi
= \\
&\frac{\alpha_2}{4\pi}
\biggl[
\left(
C_F-(Y-Q)^2
\right)X_W
+s_W^2Q^2X_\gamma
+
\frac{(Y-c_W^2Q)^2}{c_W^2}X_Z
\biggr] ,
\end{split}
\label{eq:app-general-EW}
\end{equation}
where $\alpha_{1,2}=g_{1,2}^2/(4\pi)$ and $C_F=I(I+1)$. After subtraction of the UV pole and matching the pointlike theory at the cutoff scale  $\Lambda$, the one-loop contribution $X_V$, with $V=\gamma,W,Z$, admits the heavy-mass expansion
\begin{equation}
X_V(\Lambda)
=
-2\pi m_V
+
\left[
4+3\log\!\left(\frac{\Lambda^2}{M_\chi^2}\right)
\right]  M_\chi
\ .
\end{equation}
The nonanalytic term proportional to $m_V$ is an infrared contribution and is therefore insensitive to the short-distance structure of the composite resonance. By contrast, the analytic term proportional to $M_\chi$ is sensitive to short-distance physics. 

Within a given EW multiplet, the UV-sensitive contribution proportional to $M_\chi$ is universal and therefore drops out of the mass differences. The remaining infrared contribution gives the standard charged--neutral splitting~\cite{Thomas:1998wy,Nagata:2014aoa, Cheng:1998hc, Feng:1999fu, Gherghetta:1999sw, Ibe:2012sx, McKay:2017xlc}.
\begin{equation}
\Delta_{\rm EW}\equiv M_Y^Q-M_Y^{Q=0}
=
\delta_g
\left(
Q^2+\frac{2YQ}{c_W}
\right)\ .
\label{eq:HXM_charged_splitting}
\end{equation}
with $\delta_g\equiv 
{\alpha_2}/{2}\,(1-c_W)~m_W\simeq 167\,{\rm MeV}$. Numerically, this corresponds to approximately $350\,{\rm MeV}$ for a Higgsino-like doublet and $167\,{\rm MeV}$ for a Wino-like triplet.

\smallskip
For the $\trip$ resonance, hypercharge gauging also distinguishes the neutral components of the $Y=0$ and $Y=1$ triplets. Their one-loop mass difference is
\begin{equation}
\begin{aligned}
\Delta_{10}^{\rm EW}&
\equiv
M_{Y=1}^{Q=0}-M_{Y=0}^{Q=0}=
\\
&
\frac{\alpha_1}{4\pi}
\left[
4+3\log\!\left(\frac{\Lambda^2}{M_\chi^2}\right)
\right]M_\chi
+
\frac{\alpha_2}{2}
\left(
m_W-\frac{m_Z}{c_W^2}
\right).
\end{aligned}
\label{eq:app-Delta10-EW}
\end{equation}
The second term in Eq.~\eqref{eq:app-Delta10-EW} is the universal IR contribution and is approximately $-0.65\,{\rm GeV}$. The first term instead arises from short-distance momenta in the pointlike resonance theory and is therefore sensitive to the UV completion. In the composite theory it must be supplemented by local $\SO(5)$-breaking contributions generated at the matching scale $\Lambda$. For the $\trip$ resonance, we absorb these short-distance effects into a coefficient $c\sim\mathcal{O}(1)$, which includes the pointlike hypercharge contribution $4+3\log(\Lambda^2/M_\chi^2)$ together with the contributions from local operators discussed below. For a composite resonance, these terms from higher-dimensional operators correspond to a form factor with typical length scale $\Lambda^{-1}$.

\medskip
\para{Tree-level splitting and mixing operators}
We organize the contributions to the mass spectrum according to their dimension and transformation properties under $\SO(4)\times\U(1)_X$.

\smallskip
Apart from the invariant common mass term in Eq.~\eqref{eq:app-kinetic}, no independent dimension-three mass-splitting operator exists for the real bidoublet $\bid$. In particular, EW gauge invariance forbids a Majorana mass for an individual doublet, while an insertion of $T_R^3$ vanishes once the reality condition is imposed.

\smallskip
By contrast, for the complex $\trip$ resonance, two independent dimension-three structures are relevant,
\begin{equation}
\begin{split}
\mathcal L_{(3)}
={}&
M_\chi\varepsilon_3\,
\Tr\!\left[\bar\psi\psi T_R^3\right]
+
\frac{M_\chi}{2}\slashed{\varepsilon}_3
\left[
\Tr\!\left[
\overline{\psi}\psi^c\,T_R^-
\right]
+\mathrm{h.c.}
\right],
\end{split}
\label{eq:appHXM_dim3}
\end{equation}
where $T_R^-=T_R^1-iT_R^2$. Replacing $T_R^3$ by $X$ would only induce a common mass shift within a fixed $(\mathbf n,\mathbf m)_X$ multiplet. The first term in Eq.~\eqref{eq:appHXM_dim3} preserves $\U(1)_X$ but explicitly breaks $\SU(2)_R$ to its $\U(1)_R$ subgroup. For the $\trip$, it shifts all the states of the $Y=0$ triplet relative to those of the $Y=1$ triplet, without mixing the two multiplets. For the neutral components, it provides an additional contribution to the EW splitting $\Delta^{\rm EW}_{10}$,
\begin{equation}
\Delta_{10}^{(3)}
\equiv
M_{Y=1}^{Q=0}-M_{Y=0}^{Q=0}
=
\varepsilon_3 M_\chi \ .
\label{eq:Delta10-dim3}
\end{equation}
The second term violates $\U(1)_X$ while preserving EW gauge invariance and the stabilizing parity. The two $\mathbf{3}_0$ multiplets combine into a $Y=0$ Dirac triplet, and this operator generates a Majorana mass that splits it into two real $\SU(2)_L$ triplets. Equivalently, in the neutral sector it induces maximal mixing between the two neutral $\mathbf{3}_0$ states, yielding a pseudo-Dirac pair of Majorana mass eigenstates. No analogous dimension-three Majorana mass is allowed for the $Y=1$ triplet, since it would violate hypercharge gauge invariance.

\smallskip
At dimension five we retain the $\U(1)_X$-preserving operators, since for the complex $\trip$ the leading $\U(1)_X$-violating contribution is already present at dimension three, while for the
real bidoublet $X=0$ and $\U(1)_X$ acts trivially. For the $\SU(2)_R$ doublets considered here, and assuming CP conservation, the relevant operators are~\cite{SO4neutralinos_DiLecce}
\begin{equation}
\begin{split}
\mathcal L_{(5)}
=
& \frac{g_\psi^2}{M_\chi} \,
\varepsilon_5\, \zeta
\Tr\!\left[
\bar\psi T_L^a\psi T_R^b
\right]
\Tr\!\left[
\mathcal H^\dagger
\frac{\sigma^a}{2}
\mathcal H
\frac{\sigma^b}{2}
\right] +
\\
&
\frac{g_\psi^2}{M_\chi} \,
\slashed{\varepsilon}_5\, \zeta
\Tr\!\left[
\bar\psi T_L^a\psi \slashed T_R^b
\right] 
\Tr\!\left[
\mathcal H^\dagger
\frac{\sigma^a}{2}
\mathcal H
\frac{\sigma^b}{2}
\right] \ ,
\end{split}
\label{eq:appdim5operatorsbreaking}
\end{equation}
where we define $\slashed T_R^b\equiv\sigma_3T_R^b\sigma_3$. The operator proportional to $\varepsilon_5$ is invariant under the full custodial $\SO(4)$ symmetry, whereas the one proportional to $\slashed{\varepsilon}_5$ preserves only $T_R^3$. Both preserve the SM gauge symmetry and $\U(1)_X$. In general, further $\SO(4)$-breaking dimension-5 operators can be written~\cite{SO4neutralinos_DiLecce}, that result in the same splitting among neutral states as \eqref{eq:appdim5operatorsbreaking}.
Using $\langle\mathcal H\rangle=v\mathbf1_2/\sqrt2$, $\Tr\!\left[
\mathcal H^\dagger
{\sigma^a}
\mathcal H
{\sigma^b}
\right]
=
{v^2}\delta^{ab}$, together with $\slashed T_R^{1,2}=-T_R^{1,2}$ and $\slashed T_R^3=T_R^3$, it is convenient to define
\begin{equation}
\varepsilon_{\rm spl}
=
\frac{\varepsilon_5+\slashed{\varepsilon}_5}{2},
\qquad
\varepsilon_{\rm mix}
=
\frac{\varepsilon_5-\slashed{\varepsilon}_5}{2}.
\label{eq:eps-spl-mix}
\end{equation}
The dimension-five operators in Eq.\eqref{eq:appdim5operatorsbreaking} simplify 
\begin{equation}
\mathcal L_{(5)}
=
\frac{\xi M_\chi}{2}
\bigg[
\varepsilon_{\rm spl}\,\zeta
\Tr\!\left[
\bar\psi T_L^3\psi T_R^3
\right]
+
\varepsilon_{\rm mix}\,\zeta
\sum_{a=1,2}
\Tr\!\left[
\bar\psi T_L^a\psi T_R^a
\right]
\bigg] .
\label{eq:spldim5op}
\end{equation}
Thus, $\varepsilon_{\rm spl}$ and $\varepsilon_{\rm mix}$ parameterize, respectively, the diagonal and mixing structures induced by the dimension-five operators in the resonance mass matrices. Their detailed impact depends on the decomposition of the strong resonance into EW multiplets and will be worked out explicitly below for both the $\bid$ and $\trip$ cases. In contrast to the EW radiative splittings and the dimension-three contributions, these dimension-five corrections are suppressed by $\xi$ and scale as $\xi M_\chi$.

\section{Mass spectrum and inelastic cross sections}
\label{app:spectrum-XS}
\setcounter{equation}{0}\renewcommand{\theequation}{C\arabic{equation}}
\renewcommand{\theHequation}{appendixC.\arabic{equation}}

In this Appendix, we discuss the mass spectra of the $\bid$ and $\trip$
resonances and the inelastic cross sections between the two neutral states of their light Majorana pairs.

\medskip
\para{The $\bid$ resonance}
The real bidoublet $\bid$  contains the two conjugate EW doublets $\mathbf{2}_{\pm 1/2}$, corresponding to the usual Higgsino field content. Their neutral components form a Dirac fermion before the inclusion of symmetry-breaking mass corrections. As discussed above, the leading local mass corrections for the bidoublet arise at dimension five, since the dimension-three operators in Eq.~\eqref{eq:appHXM_dim3} vanish for the real representation. 

Applying Eq.~\eqref{eq:spldim5op} to the $\bid$, the combinations proportional to $\varepsilon_{\rm spl}$ and $\varepsilon_{\rm mix}$ correspond, respectively, to the operators $\mathcal O_+$ and $\mathcal O_0$ of Ref.~\cite{Bottaro:2022one}. The former contributes, together with the EW correction in Eq.~\eqref{eq:HXM_charged_splitting}, to the charged--neutral mass splitting. In particular, the contribution from ${\cal O}_+$ can be significant when $\xi M_\chi \varepsilon_{\rm spl}/4$ is comparable to the Higgsino EW charged-neutral mass splitting of about $350~\rm MeV$. However, its precise value is irrelevant for the following discussion because one can always change the sign of $\varepsilon_{\rm spl}$ in order to preserve the right mass ordering.

The latter instead breaks the accidental fermion-number symmetry of the neutral Dirac pair and generates its Majorana splitting. The resulting matrix of mass corrections is
\begin{equation}
\Delta {\mathbf M}
=
\frac{\xi M_\chi}{8}
\begin{pmatrix}
\varepsilon_{\rm spl} & 2\varepsilon_{\rm mix}\\
2\varepsilon_{\rm mix} & \varepsilon_{\rm spl}
\end{pmatrix} \ .
\label{eq: Appendix mass matrix 5}
\end{equation}
The equal diagonal corrections imply maximal mixing between the two neutral components. After diagonalization, the two Majorana mass eigenstates are denoted by $\chi_0$ and $\chi_1$. The contribution proportional to $\varepsilon_{\rm spl}$ shifts their common mass and therefore drops out of the Majorana splitting, which is entirely controlled by $\varepsilon_{\rm mix}$:
\begin{equation}
\delta
\equiv
m_{\chi_1}-m_{\chi_0}
=
\frac{\xi M_\chi}{2}
|\varepsilon_{\rm mix}| \ .
\label{eq:HXM_delta22}
\end{equation}
This is the Majorana splitting introduced in Eq.~\eqref{eq:delta22}, which controls the kinematics of inelastic scattering. Along the thermal relic trajectory in the strong-annihilation regime, $\xi M_\chi$ is approximately constant, so that a fixed $\varepsilon_{\rm mix}$ corresponds to an approximately mass-independent inelastic splitting.

Using the neutral current derived in Appendix~\ref{app:currents}, the $Z$ interaction becomes purely off-diagonal in the Majorana mass basis, thereby mediating the inelastic transition $\chi_0\to\chi_1$:
\begin{equation}
\mathcal L^Z_{\bid}
=
i\frac{g_2}{2c_W}\left(1-\frac{\xi}{2}\right)\,
Z_\mu\bar\chi_0\gamma^\mu\chi_1,
\label{eq:app-bid-Z}
\end{equation}
where the leading term is the standard Higgsino coupling, while the $\mathcal O(\xi)$ correction is induced by the composite current in Eq.~\eqref{eq:strong}. The corresponding zero-momentum inelastic cross section per neutron is
\begin{equation}
\sigma_n
=
\sigma_n^{\rm EW}
\left(1-\frac{\xi}{2}\right)^2 \ ,
\label{eq:HXM_sigma22}
\end{equation}
where $\sigma_n^{\rm EW} \equiv
G_F^2\mu_{\chi n}^2/(2\pi) \simeq
7.43\times10^{-39}\,{\rm cm}^2$ for $M_\chi \gg m_n$. 

\medskip
\para{The $\trip$ resonance}
The complex $(\mathbf3,\mathbf2)_{1/2}$ resonance together with its conjugate contains two real $\mathbf3_0$ multiplets and the conjugate pair $\mathbf3_{\pm1}$. We denote the corresponding neutral Dirac states by $\psi_0^0$ and $\psi_1^0$, respectively.

Applying the dimension-five mass corrections in Eq.~\eqref{eq:spldim5op} to the $\trip$ and including the EW and dimension-three contributions to the splitting between $\psi_0^0$ and $\psi_1^0$, the neutral spectrum can first be described in the $\U(1)_X$-preserving limit. The resulting matrix of mass corrections is
\begin{equation}
\Delta {\mathbf M}
=
\begin{pmatrix}
-\Delta_{10}/2 & -\sqrt{2} \,\mu_{10}\\
-\sqrt{2} \,\mu_{10} & \Delta_{10}/2
\end{pmatrix} \ , \qquad
\mu_{10} = \frac{\xi M_\chi}{4}\varepsilon_{\rm mix} \ .
\label{eq:HXM_Lmix}
\end{equation}
The quantity $\Delta_{10}$ is the total mass difference between the neutral components of the $Y=1$ and $Y=0$ Dirac triplets. It receives the EW contribution in Eq.~\eqref{eq:app-Delta10-EW}, the dimension-three symmetry-breaking contribution in Eq.~\eqref{eq:Delta10-dim3}, and the diagonal dimension-five correction proportional to $\varepsilon_{\rm spl}$, yielding
\begin{equation}
\begin{split}
\Delta_{10}
\equiv
\Delta_{10}^{\rm EW}
+
\Delta_{10}^{(3)}
-
\frac{\xi M_\chi}{4}\varepsilon_{\rm spl}
=
c\,\frac{\alpha_1}{4\pi}M_\chi
+
\frac{\alpha_2}{2}
\left(
m_W-\frac{m_Z}{c_W^2}
\right) .
\end{split}
\end{equation}
The coefficient $c$ combines the short-distance hypercharge contribution computed in the pointlike theory with the local $\SO(5)$-breaking contributions. Explicitly,
\begin{equation}
c = \bigg[4 +
3\log\!\left(\frac{\Lambda^2}{M_\chi^2}\right)\bigg]
+
4\pi\bigg[\,
\frac{\varepsilon_3}{\alpha_1}
-
\frac{\xi\,\varepsilon_{\rm spl}}{4\alpha_1}
\bigg]\ .
\end{equation}
As discussed in the main text and Appendix~\ref{app:recast}, this coefficient, first introduced in Eq.~\eqref{eq:Delta10-main}, is treated as a free parameter. In summary, $\varepsilon_{\rm spl}$ contributes to the diagonal separation between the two neutral Dirac states, whereas $\varepsilon_{\rm mix}$ induces their mixing. 

For real coefficients, diagonalizing the neutral Dirac mass matrix gives the light and heavy eigenstates
\begin{equation}
\psi_\ell^0
=
\cos\theta\,\psi_0^0-\sin\theta\,\psi_1^0,
\qquad
\psi_h^0
=
\sin\theta\,\psi_0^0+\cos\theta\,\psi_1^0,
\label{eq:HXM_neutral_rotation}
\end{equation}
with mixing angle
\begin{equation}
\tan(2\theta)
=
-\frac{\xi M_\chi\varepsilon_{\rm mix}}
{\sqrt2\,\Delta_{10}} \ .
\label{eq:HXM_theta32}
\end{equation}
This is the mixing angle introduced in the main text in
Eq.~\eqref{eq:theta32}. It controls the admixture of the hypercharged $Y=1$ component in the light state and therefore plays a central role in determining its coupling to the $Z$ boson and the corresponding inelastic direct-detection rate. We focus on the branch $\Delta_{10}>0$, for which the light state is predominantly $Y=0$ and $|\theta|<\pi/4$. The opposite ordering instead gives a predominantly hypercharged light state, whose neutral current is dominated by the standard EW interaction. For $|\varepsilon_{\rm spl}|\lesssim0.5$, its contribution to $\Delta_{10}$ remains subleading in the parameter region of interest, since it is suppressed by $\xi$. The dimension-five interactions preserve $\U(1)_X$ and therefore leave $\psi_\ell^0$ and $\psi_h^0$ as Dirac fermions.
The same mass corrections also affect the charged spectrum. For the hypercharged states, the dominant correction is $\Delta_{10}$, which raises the entire triplet relative to the $Y=0$ one. The $Q=\pm1$ states of the $Y=0$ triplet receive a positive infrared EW shift relative to the neutral component, along with a strong-sector correction. As in the Higgsino case, the strong-sector coefficients can be chosen so that the lightest state is neutral. In this case, several operators contribute to the neutral and charged mass splittings, and can be adjusted independently~\cite{SO4neutralinos_DiLecce}.

The Majorana splitting is instead generated by the $\U(1)_X$-violating contribution discussed above and parametrized by $\slashed{\varepsilon}_3$ in Eq.~\eqref{eq:appHXM_dim3}. Since this interaction acts only on the $\mathbf3_0$ component, its projection onto the light Dirac state is suppressed by its $\mathbf3_0$ fraction, $\cos^2\theta$. In analogy with the $\bid$ case, the light Dirac state is therefore split into a
pseudo-Dirac pair of Majorana mass eigenstates, $\chi_0$ and $\chi_1$. To leading order in $\U(1)_X$ violation, their mass splitting is
\begin{equation}
\delta
\equiv
m_{\chi_1}-m_{\chi_0}
=
2|\slashed{\varepsilon}_3|M_\chi\cos^2\theta .
\label{eq:HXM_delta32}
\end{equation}
Thus, the mixing angle $\theta$ controls the hypercharged admixture of the light state, whereas $\slashed{\varepsilon}_3$ controls its inelastic Majorana splitting.

Using the neutral current derived in Appendix~\ref{app:currents}, the $Z$ interaction of the light Majorana pair is
\begin{equation}
\mathcal L^Z_{\trip}
=
-i\frac{g_2}{c_W}
\left[
\sin^2\theta
+
\frac{\xi}{8}
\left(
1-4\sin^2\theta
\right)
\right]
Z_\mu\bar\chi_0\gamma^\mu\chi_1 .
\label{eq:app-light-inelastic-current}
\end{equation}
The first term is inherited from the hypercharged component of the light state and is therefore controlled by the mixing angle $\theta$. The second originates from the composite current and remains nonzero in the pure $Y=0$ limit. The corresponding zero-momentum inelastic cross section per neutron is
\begin{equation}
\sigma_n
=
\sigma_n^{\rm EW}
\left[
2\sin^2\theta
+
\frac{\xi}{4}
\left(
1-4\sin^2\theta
\right)
\right]^2,
\label{eq:HXM_sigma32}
\end{equation}
Thus, unlike the Higgsino-like case, the transition strength is not fixed at the EW value: it is controlled by the hypercharged admixture of the light state, with a residual $\mathcal O(\xi)$ interaction surviving in the $Y=0$ limit.

\section{Halo model, recoil rate and likelihood}\label{app:recast}
\setcounter{equation}{0}\renewcommand{\theequation}{D\arabic{equation}}
\renewcommand{\theHequation}{appendixD.\arabic{equation}}
\renewcommand{\thefigure}{D\arabic{figure}}
\renewcommand{\theHfigure}{appendixD.\arabic{figure}}
In this Appendix, we provide details of the halo model, the recoil rate calculation, and the statistical analysis employed.

\medskip
\para{The DM Halo model}
We use the isotope-resolved recoil rate in Eq.~\eqref{eq:isotope}, with
$\mathcal N_A$ fixed by the natural xenon abundances~\cite{Meija:2016}.
Equivalently, for isotope mass fractions $w_A$, one has
$\mathcal N_A m_A=w_A$ in consistent units. The Standard Halo Model is
described by an isotropic Maxwell--Boltzmann Galactic velocity
distribution truncated at $v_{\rm esc}$~\cite{Baxter:2021pqo}.
We adopt $\rho_\chi=0.3~\mathrm{GeV\,cm^{-3}}$, $v_0=238~\mathrm{km\,s^{-1}}$,
and $v_{\rm esc}=544~\mathrm{km\,s^{-1}}$, together with
$\mathbf v_{\rm LSR}=(0,238,0)~\mathrm{km\,s^{-1}}$ and
$\mathbf v_{\rm pec}=(11.1,12.2,7.3)~\mathrm{km\,s^{-1}}$
~\cite{Schoenrich:2009bx}. The Earth's orbital speed is taken to be
$29.8~\mathrm{km\,s^{-1}}$~\cite{Baxter:2021pqo}, and its motion
is included explicitly in the laboratory boost. For a Galactic
distribution $f_{\rm Gal}$ and laboratory velocity
$\mathbf v_{\rm lab}(t)$, we use
$f_{\rm lab}(\mathbf v,t)=f_{\rm Gal}(\mathbf v+\mathbf v_{\rm lab}(t))$.
The mean velocity integral is
\begin{equation}
\begin{split}
\eta(v_{\min},t)
&=
\int_{|\mathbf v|>v_{\min}} d^3v\,
\frac{f_{\rm lab}(\mathbf v,t)}{|\mathbf v|} \ ,
\\
\overline\eta(v_{\min})
&=
\frac{1}{T}\int_0^T dt\,\eta(v_{\min},t) \ ,
\label{eq:eta}
\end{split}
\end{equation}
where $T$ is one sidereal year. Annual averaging is relevant here because
inelastic scattering can probe the high-velocity tail of the halo
distribution. The minimum of Eq.~\eqref{eq:vmin} occurs at
\begin{equation}
E_R^\star=\frac{\mu_{\chi A}}{m_A}\delta \ ,
\qquad
v_{\min}^\star=\sqrt{\frac{2\delta}{\mu_{\chi A}}} \ ,
\end{equation}
so that decreasing $E_R$ below $E_R^\star$ increases the minimum
incident velocity and suppresses the low-energy recoil spectrum.

\medskip
\para{Recoil rate calculations}
At the recoil energies relevant for LZ230616, the momentum transfer is
of order a few hundred MeV and the loss of nuclear coherence is
important. We therefore use the Vietze isotope-dependent spin-independent
structure factors~\cite{Vietze:2014blah} for
$^{128,129,130,131,132,134,136}$Xe, normalized to unity at zero momentum
transfer and evaluated at $q=\sqrt{2m_AE_R}$. The additional one-body
contribution for $^{131}$Xe is included. We also verified the robustness of our results against the choice of nuclear form factor by adopting the Helm parametrization~\cite{Lewin:1995rx,Helm:1956zz}, as discussed in
Appendix~\ref{app:robustness}.

We identify two disjoint regions: the LZ region of interest (ROI), which
contains the event at $E_{\rm obs}=248~\mathrm{keV}$, and the
empty high-energy sideband (HSB). The LZ ROI is defined in corrected
scintillation and ionization coordinates by
$3<S1c<600$ phd, $S2>645$ phd, and
$10^{2.75}<S2c<10^{4.15}$ phd~\cite{LZ:2026axp}, corresponding to
nuclear-recoil energies up to approximately $270~\mathrm{keV}$. We
digitize the nuclear-recoil efficiency $\epsilon_{\rm LZ}(E_R)$ from
Fig.~S2 of Ref.~\cite{LZ:2026axp} and use an exposure
$\mathcal E=2.84~\mathrm{tonne\,yr}$.

To incorporate the measured recoil energy, we convolve the accepted
spectrum in the ROI with a normalized Gaussian response,
\begin{equation}
G(E_{\rm rec},E_R)
=
\frac{1}{\sqrt{2\pi}\sigma_E(E_R)}
\exp\left[
-\frac{(E_{\rm rec}-E_R)^2}{2\sigma_E^2(E_R)}
\right] \ ,
\end{equation}
with
$\sigma_E(E_R)
=
\sqrt{
(23~\mathrm{keV})^2 E_R/(248~\mathrm{keV})
+(0.093E_R)^2
}.$
The reconstructed-energy intensity is then
\begin{equation}
\lambda_{\rm ROI}(E_{\rm rec};\boldsymbol{\varphi})
=
\mathcal E
\int dE_R\,
\epsilon_{\rm LZ}(E_R)
\frac{dR}{dE_R}
G(E_{\rm rec},E_R) \ ,
\label{eq:intensity}
\end{equation}
where the model parameters $\mathbf{\varphi}$ are understood as arguments of the rate. The corresponding expected ROI yield is
\begin{equation}
N_{\rm ROI}(\boldsymbol{\varphi})
=
\int_{\rm ROI} dE_{\rm rec}\,
\lambda_{\rm ROI}(E_{\rm rec};\boldsymbol{\varphi}) \ .
\label{eq:NROI}
\end{equation}
The HSB follows Ref.~\cite{Rodd:2026tyn} and is defined by
$800<S1c<1700$ phd and $10^{2.75}<S2c<10^{4.3}$ phd. We take it to
correspond to the recoil-energy interval
$350\lesssim E_R/\mathrm{keV}\lesssim675$. Since the nuclear-recoil
acceptance in this region is not publicly available, we assume unit
acceptance and neglect energy smearing, defining
\begin{equation}
N_{\rm HSB}(\boldsymbol{\varphi})
=
\mathcal E
\int_{\rm HSB} dE_R\,
\frac{dR}{dE_R} \ .
\label{eq:NHSB}
\end{equation}
We thus define the total expected signal yield as the sum of the contributions from the two disjoint regions,
\begin{equation}
N_{\rm sig}(\boldsymbol{\varphi})
=N_{\rm ROI}(\boldsymbol{\varphi})
+N_{\rm HSB}(\boldsymbol{\varphi}) \ .
\label{eq:Nsig}
\end{equation}
For the $\bid$ representation, the parameter space is spanned by $\boldsymbol{\varphi}=(M_\chi,\delta)$, while the mixed $\trip$ involves the additional parameters $c$ and $\varepsilon_{\rm mix}$, such that $\boldsymbol{\varphi}=(M_\chi,\delta,\varepsilon_{\rm mix},c)$. 

In Fig.~\ref{fig:dNdER} we compare the underlying recoil spectra with those including the ROI detector response for three representative benchmarks. The composite Higgsino $\mathbf{(2,2)}_0$ is shown for $M_\chi=50~\mathrm{TeV}$ and $\delta=380~\mathrm{keV}$; the zero-mixing $\mathbf{(3,2)}_{1/2}$ limit for
$M_\chi=3~\mathrm{TeV}$ and $\delta=180~\mathrm{keV}$; and the mixed $\mathbf{(3,2)}_{1/2}$ state for $M_\chi=5~\mathrm{TeV}$, $\delta=350~\mathrm{keV}$, and $|\theta|=10^\circ$. As discussed in the main text, the zero-mixing scenario predicts a recoil spectrum sharply peaked at low energies, disfavoring the observed event at $248~\mathrm{keV}$ and suggesting that a comprehensive assessment of this scenario should also incorporate the constraints from the LZ low-energy dataset. By contrast, a mixing angle of $\approx 10^\circ$ yields a spectrum resembling that of the $\bid$, but with a suppressed high-energy shoulder, reducing the expected event yield in the empty HSB.
\begin{figure}[t]
\centering
\includegraphics[width=0.95\columnwidth]{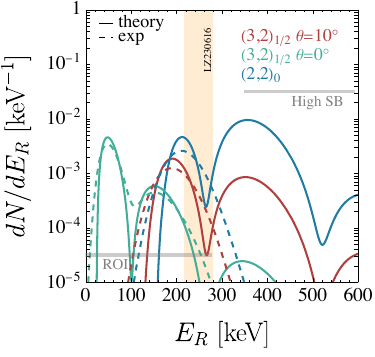}
\caption{
Differential nuclear recoil spectra for three representative composite dark-matter benchmarks. The blue curve corresponds to the composite Higgsino $\mathbf{(2,2)}_0$ with $M_\chi=50~\mathrm{TeV}$ and $\delta=380~\mathrm{keV}$. The green curve represents the pure $\mathbf{3}_0$ limit of the $\mathbf{(3,2)}_{1/2}$ multiplet, with $M_\chi=3~\mathrm{TeV}$ and $\delta=180~\mathrm{keV}$. The red curve shows the mixed $\mathbf{(3,2)}_{1/2}$ scenario with $M_\chi=5~\mathrm{TeV}$, $\delta=350~\mathrm{keV}$, and $|\theta|=10^\circ$. Solid lines indicate the underlying theoretical recoil spectra, while dashed lines represent the detector-level predictions after accounting for the recoil-energy acceptance and energy resolution. All spectra are computed using the isotope-dependent Vietze nuclear response. The highlighted interval marks the observed LZ230616 recoil at $E_{\rm obs}=248~\mathrm{keV}$, with the ROI and HSB indicated schematically. The absence of observed HSB events is incorporated into the likelihood through the corresponding expected signal yield, under the detector-response assumptions described in the text.
}
\label{fig:dNdER}
\end{figure}
Assuming negligible backgrounds in both regions, the ROI contributes through the differential reconstructed-energy rate $\lambda_{\rm ROI}(E_{\rm rec};\boldsymbol{\varphi})$,
while the empty HSB enters through its integrated expected yield. With a single observed event at $E_{\rm obs}=248~\mathrm{keV}$, the extended
unbinned log-likelihood reads as in Eq.~\eqref{eq:lz_likelihood}.
The first term accounts for the expected event counts in both regions, penalizing models that predict a sizable signal in the empty HSB, while the second incorporates the spectral information from the observed recoil. 

\bigskip
\para{Statistical Analysis}
To determine the two-dimensional allowed regions in the $(M_\chi,\delta)$ plane, we employ a Monte Carlo-calibrated likelihood-ratio test. At each tested point $\boldsymbol{\varphi}_0=(M_\chi,\delta)$, the test statistic is defined as
\begin{equation}
q(\boldsymbol{\varphi}_0)
=-2\log\frac{\mathcal L(\boldsymbol{\varphi}_0)}
{\mathcal L(\widehat{\boldsymbol{\varphi}})}
=2\left[\log\mathcal L(\widehat{\boldsymbol{\varphi}})
-\log\mathcal L(\boldsymbol{\varphi}_0)\right] \ ,
\label{eq:qstat}
\end{equation}
where $\widehat{\boldsymbol{\varphi}}$ maximizes the likelihood for
the data set being fitted. For each hypothesis tested in the
$(M_\chi,\delta)$ plane, pseudo-experiments are generated with
$N_{\rm toy}\sim\mathrm{Poisson}\!\left[
N_{\rm sig}(\boldsymbol{\varphi}_0)\right].$
Each generated event is assigned either to the ROI or to the HSB with
probabilities $N_{\rm ROI}/N_{\rm sig}$ and
$N_{\rm HSB}/N_{\rm sig}$, respectively. ROI recoil energies are
then sampled from the normalized reconstructed-energy spectrum
$\lambda_{\rm ROI}(E_{\rm rec})/N_{\rm ROI}$; the HSB contributes
only its event count. Each toy is independently refitted over the
specified benchmark parameter space to obtain its likelihood-ratio
statistic. The critical value $q_{\rm crit}^{\,C}$ is defined by
$\Pr\!\left[q_{\rm toy}(\boldsymbol{\varphi}_0)
\le q_{\rm crit}^{\,C}(\boldsymbol{\varphi}_0)
\mid\boldsymbol{\varphi}_0\right]=C,$
with $C=0.683$ and $0.95$. A point is retained when
$q_{\rm obs}(\boldsymbol{\varphi}_0)\le
q_{\rm crit}^{\,C}(\boldsymbol{\varphi}_0)$.
This is a Neyman construction with likelihood-ratio ordering
~\cite{Neyman1937,FeldmanCousins}; its critical values are not
assumed to follow the asymptotic Wilks distribution
~\cite{Wilks1938,Cowan:2010js}. The non asymptotic behavior is especially transparent for toys with no
observed events. Their likelihood contains only the term
$-N_{\rm sig}$, giving
$q_0(\boldsymbol{\varphi}_0)
=2\left[N_{\rm sig}(\boldsymbol{\varphi}_0)
-N_{\rm sig}^{\rm min}\right]
\simeq 2N_{\rm sig}(\boldsymbol{\varphi}_0),$
when the minimum accessible yield is close to zero. This branch has
probability $\exp[-N_{\rm sig}(\boldsymbol{\varphi}_0)]$ and can
strongly affect the calibrated quantiles. For example, for the
low-rate limiting solution with
$N_{\rm sig}=0.325$, the zero-event probability is approximately
$0.72$. Thus even the $68.3\%$ critical value can be determined by
the discrete zero-event branch rather than by an asymptotic $\chi^2$ approximation. The dark and light blue contours in
Fig.~\ref{fig:results} and in the right panel of Fig.~\ref{fig:posterior} refer exclusively
to this calibrated frequentist construction.

\medskip
To assess which mixing configurations are favored by the recoil
event, we additionally construct a Bayesian posterior for the
mixed $\mathbf{(3,2)}_{1/2}$ model. The mixing angle $\theta=\theta(M_\chi,\varepsilon_{\rm mix},c)$ is determined by diagonalizing the neutral-state mass matrix, as described in Eq.~\eqref{eq:theta32}. The resulting event yields and reconstructed-energy spectrum enter the same extended unbinned likelihood defined in Eq.~\eqref{eq:lz_likelihood}. We restrict the parameter space to the branch in which the predominantly $Y=0$ state is the lightest neutral Dirac eigenstate, requiring a positive mass-ordering parameter $\Delta_{10}(M_\chi,c)>0$. In the opposite configuration, the phenomenology is mostly driven by EW interactions, as in the $\bid$ case.

\begin{figure*}[t]
\centering
\includegraphics[width=0.49\textwidth]{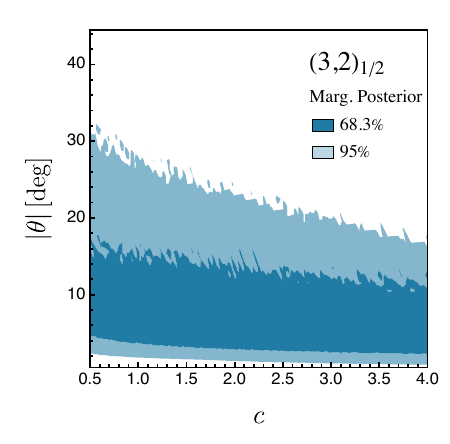}%
\hfill
\includegraphics[width=0.49\textwidth]{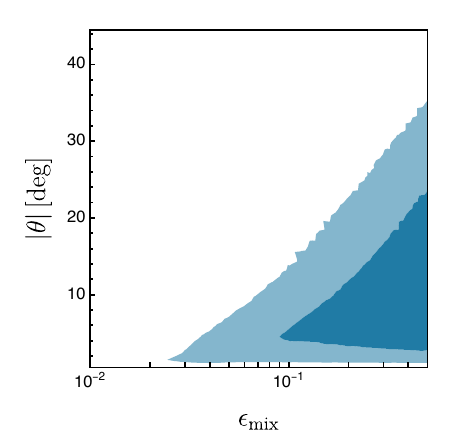}%
\caption{
Marginalized Bayesian posterior distributions in the $(c,|\theta|)$ (left) and $(\varepsilon_{\rm mix},|\theta|)$ (right) planes for the mixed $\mathbf{(3,2)}_{1/2}$ model, inferred from the LZ230616 recoil event. Dark and light blue regions denote the $68.3\%$ and $95\%$ HPD regions, respectively, obtained by marginalizing over the remaining model parameters. The analysis incorporates the empty high-energy sideband and imposes the mass-ordering condition $\Delta_{10}>0$.
}
\label{fig:posteriorApp}
\end{figure*}

Denoting by $D$ the observed recoil at $248~\mathrm{keV}$, the normalized posterior reads,
\begin{equation}
p(\boldsymbol{\varphi}\mid D)
=\frac{\mathcal L(D\mid\boldsymbol{\varphi})
\,\pi(\boldsymbol{\varphi})\,
}
{\displaystyle\int d\boldsymbol{\varphi}'\,
\mathcal L(D\mid\boldsymbol{\varphi}')
\pi(\boldsymbol{\varphi}')\,
} \ ,
\label{eq:posterior4d}
\end{equation}
where $\boldsymbol{\varphi}=(\log_{10}(M_\chi/\mathrm{TeV}),\delta,\varepsilon_{\rm mix},c)$ and $\pi(\varphi)$ denotes the prior restricted to the physically allowed parameter space. For the posterior projections in the left panel of 
Fig.~\ref{fig:posterior} and in Fig.~\ref{fig:posteriorApp}, we use
$2\le M_\chi/\mathrm{TeV}\le200$,
$100\le\delta/\mathrm{keV}\le500$,
$10^{-4}\le\varepsilon_{\rm mix}\le0.5$, and
$0.5\le c\le4$. The reference prior is uniform in
$\log_{10}(M_\chi/\mathrm{TeV})$, $\delta$,
$\varepsilon_{\rm mix}$, and $c$ before the mass-ordering cut.
The constraint $\Delta_{10}>0$ is imposed at the level
of the joint prior, with one overall normalization; we do not
renormalize the allowed $c$ interval separately at each mass.
Consequently, masses with different physically allowed volumes
need not carry the same marginal prior weight. The highest-posterior-density (HPD) regions
are defined with respect to the plotting variables and
their associated integration measures.

Writing $\vartheta\equiv|\theta|$ for the absolute mixing angle, without introducing any additional physical parameter, the one-dimensional mixing-angle posterior is obtained by marginalizing the joint posterior over the remaining parameters, while accounting for the dependence of $\vartheta$ on the model parameters,
\begin{equation}
p(\vartheta\mid D)=\int d\boldsymbol{\varphi}\,
 p(\boldsymbol{\varphi}\mid D)\,
\delta_{\rm D}\!\left(\vartheta-|\theta(\boldsymbol{\varphi})|\right) \ ,
\label{eq:theta_posterior}
\end{equation}
where $\vartheta\in[0,45^\circ]$. We also construct the prior-induced distribution using the same procedure without the likelihood factor. Although a detailed comparison of these distributions lies beyond the scope of this work and is therefore not shown, we have verified that the posterior exhibits appreciable support for mixing angles around $10^\circ$, in contrast to the strong preference for near-zero mixing induced by the prior measure. 

The posterior projections in the left panel of Fig.~\ref{fig:posterior} and in Fig.~\ref{fig:posteriorApp} are obtained by
marginalizing the joint posterior over the remaining parameters, treating $|\theta|$ as a derived quantity fixed by $(M_\chi,\varepsilon_{\rm mix},c)$.
The $68.3\%$ and $95\%$ highest-posterior-density (HPD) regions enclose the corresponding integrated probabilities in the displayed variables. 

The upper plots in the left panel of Fig.~\ref{fig:posterior} reveal an anticorrelation between $M_\chi$ and $|\theta|$ and a positive correlation between $\delta$ and $|\theta|$. Larger mixing angles are favored at lower masses and higher splittings, reflecting the competition between the inelastic kinematic suppression and the transition coupling. Increasing $\delta$ probes the high-velocity tail of the Galactic halo, suppressing the scattering rate, which can be compensated by stronger mixing until the kinematic boundary is reached. The complementary projections in Fig.~\ref{fig:posteriorApp} show that appreciable mixing requires sufficiently large $\varepsilon_{\rm mix}$, while the upper envelope of $|\theta|$ decreases with increasing $c$.

Finally, marginalizing over both mixing parameters yields the mass--splitting posterior shown in the lower-left plot of the left panel of Fig.~\ref{fig:posterior},
\begin{equation}
p(\log_{10}M_\chi,\delta\mid D)
=\int d\varepsilon_{\rm mix}\,dc\,
 p(\log_{10}M_\chi,\delta,
 \varepsilon_{\rm mix},c\mid D) \ ,
\label{eq:md_posterior}
\end{equation}
whose highest posterior densities occur at $M_\chi$ of a few to $\sim10$--$15~\mathrm{TeV}$ and $\delta\simeq250$--$390~\mathrm{keV}$. The $95\%$ HPD region extends to several tens of TeV and $\delta\simeq170$--$400~\mathrm{keV}$. The shape of the marginalized posterior is driven by both the recoil-spectrum likelihood and the prior-weighted volume of physically allowed mixing configurations.

\section{Benchmarks and robustness}
\label{app:robustness}
\setcounter{equation}{0}\renewcommand{\theequation}{E\arabic{equation}}
\renewcommand{\theHequation}{appendixE.\arabic{equation}}
\setcounter{table}{0}
\renewcommand{\thetable}{E\arabic{table}}
\renewcommand{\theHtable}{appendixE.\arabic{table}}

\begin{table}[t]
\centering
\caption{
Mass splittings yielding $N_{\rm ROI}=1$ for representative
DM benchmarks and different Galactic escape
velocities. We compare the isotope-dependent nuclear
response of Ref.~\cite{Vietze:2014blah} with the Helm
form factor. The high-energy sideband yield
$N_{\rm HSB}$ is evaluated using the Vietze response
at the corresponding Vietze splitting, assuming
unit sideband acceptance. The mixed benchmarks
correspond to fixed mixing angles of
$|\theta|=10^\circ$ and $40^\circ$.
}
\label{tab:benchmarks}
\small
\begin{tabular}{@{}llrrrrr@{}}
\toprule
Multiplet & DM & $M_\chi$ & $v_{\rm esc}$
& $\delta^{\rm Vietze}$ & $\delta^{\rm Helm}$
& $N_{\rm HSB}^{\rm Vietze}$\\
& & [TeV] & [km/s] & [keV] & [keV] & \\
\midrule

$\trip$ & $\mathbf{3}_0$
& 4 & 544 & 127.1 & 129.0 & $\ll1$\\
& & & 567 & 127.7 & 129.7 & $\ll1$\\
& & & 610 & 128.6 & 130.6 & $\ll1$\\

\midrule

$\bid$ & $\mathbf{2}_{1/2}$
& 50 & 544 & 366.8 & 377.0 & 1.3\\
& & & 567 & 379.3 & 390.5 & 1.3\\
& & & 610 & 400.5 & 414.0 & 1.3\\

\midrule

$\trip$ & $|\theta|=10^\circ$
& 5 & 544 & 314.1 & 332.3 & 0.2\\
& & & 567 & 321.3 & 341.3 & 0.2\\
& & & 610 & 331.6 & 355.2 & 0.2\\

\midrule

$\trip$ & $|\theta|=40^\circ$
& 5 & 544 & 378.6 & 386.6 & 3.5\\
& & & 567 & 393.0 & 401.8 & 3.7\\
& & & 610 & 418.3 & 428.8 & 3.7\\

\bottomrule
\end{tabular}
\end{table}

Table~\ref{tab:benchmarks} summarizes the mass splittings
required to obtain $N_{\rm ROI}=1$ for representative
DM benchmarks, together with their predicted
high-energy sideband populations. We consider three values
of the Galactic escape velocity,
$v_{\rm esc}=\{544,567,610\}~\mathrm{km\,s^{-1}}$,
and compare the isotope-dependent nuclear response of
Ref.~\cite{Vietze:2014blah} with the Helm form factor~\cite{Helm:1956zz}.
The sideband yields are evaluated at the corresponding
Vietze splitting, assuming unit HSB acceptance.
These comparisons quantify the dependence on astrophysical
and nuclear inputs before applying complementary constraints
from solar capture and indirect detection.

The sensitivity to the Galactic escape velocity varies
significantly across the benchmarks. In the pure Wino-like limit at $M_\chi=4~\mathrm{TeV}$,
increasing $v_{\rm esc}$ from $544$ to
$610~\mathrm{km\,s^{-1}}$ shifts the required splitting
from $127.1$ to $128.6~\mathrm{keV}$, corresponding
to a variation of approximately $1\%$.
The predicted HSB population remains negligible,
$N_{\rm HSB}\sim10^{-3}$, throughout the considered
range. This mild dependence reflects the relatively
small inelastic threshold, for which the recoil spectrum
is less sensitive to the extreme high-velocity tail of
the halo distribution.

By contrast, the composite Higgsino at
$M_\chi=50~\mathrm{TeV}$ requires significantly
larger splittings to suppress its unsuppressed
EW transition rate. The splitting increases
from $366.8$ to $400.5~\mathrm{keV}$ as
$v_{\rm esc}$ varies from $544$ to
$610~\mathrm{km\,s^{-1}}$, a shift of approximately
$9\%$. The corresponding sideband population remains
of order unity, $N_{\rm HSB}\simeq1.3$.
This stronger dependence arises because the large
inelastic threshold forces the scattering process
to probe the high-velocity tail of the halo
distribution.
The mixed $\trip$ benchmarks illustrate the impact of the neutral-state mixing angle on the recoil predictions. At fixed DM mass $M_\chi=5\,\mathrm{TeV}$, the inelastic transition strength depends on $\theta$ through Eq.~\eqref{eq:sigma32}. For moderate mixing, $|\theta|=10^\circ$, the suppressed transition current allows one ROI event to be reproduced with splittings between $314.1$ and $331.6\,\mathrm{keV}$ as $v_{\rm esc}$ increases from $544$ to $610\,\mathrm{km\,s^{-1}}$. The corresponding variation is approximately $5.6\%$, intermediate between the pure Wino-like and composite-Higgsino limits. The predicted HSB yield remains small, $N_{\rm HSB}\simeq0.18$--$0.19$, corresponding to a probability of approximately $82$--$84\%$ of observing zero sideband events.

For larger mixing, $|\theta|=40^\circ$, the enhanced EW transition current requires a greater splitting to maintain $N_{\rm ROI}=1$, yielding $\delta=378.6$--$418.3\,\mathrm{keV}$ across the same range of escape velocities. The resulting variation is approximately $10.5\%$, while the predicted sideband population increases significantly to $N_{\rm HSB}\simeq3.5$--$3.7$. Consequently, the probability of observing zero HSB events decreases to approximately $2.4$--$3.0\%$.

Although all benchmarks are normalized to reproduce one event in the ROI, their predicted sideband populations differ substantially. Increasing the mixing angle enhances the transition rate, requiring a larger inelastic splitting to compensate for the stronger coupling. The resulting change in the recoil spectrum can nevertheless leave a significant population at higher energies. The absence of HSB events therefore provides an additional constraint on large-mixing configurations, illustrating the importance of incorporating both the observed recoil and the empty sideband in the likelihood analysis.

The dependence on the nuclear response is also
nontrivial. For the $|\theta|=10^\circ$ benchmark,
replacing the isotope-dependent Vietze response
with the Helm form factor increases the splitting
required to reproduce one ROI event by approximately
$18$--$24~\mathrm{keV}$, depending on
$v_{\rm esc}$. For $|\theta|=40^\circ$, the
corresponding shifts are smaller,
approximately $8$--$11~\mathrm{keV}$.
The nuclear-response uncertainty therefore
affects the moderate-mixing benchmark more strongly
than the near-maximal-mixing benchmark in terms
of the splitting required to maintain the
same ROI normalization.
Overall, these comparisons show that the mixing angle controls the normalization of the inelastic scattering rate and, through the requirement $N_{\rm ROI}=1$, indirectly selects different recoil spectra by shifting the preferred mass splitting. Moderate mixing can accommodate the observed recoil with relatively few HSB events, whereas configurations approaching maximal mixing can predict a substantial sideband population. 

Throughout the main text and in Fig.~\ref{fig:results} and the right panel of Fig.~\ref{fig:posterior}, we adopt $v_{\rm esc}=544~\mathrm{km\,s^{-1}}$ and the isotope-dependent Vietze nuclear response as our reference configuration.

\section{Solar capture}
\label{app:sc}
\setcounter{equation}{0}\renewcommand{\theequation}{F\arabic{equation}}
\renewcommand{\theHequation}{appendixF.\arabic{equation}}
The gray shaded regions in Figs.~\ref{fig:results} and \ref{fig:posterior} are obtained as follows. For the $\bid$, we adopt the constraint represented by the black curve in the right panel of Fig.~2 of Ref.~\cite{Luengas:2026tlf}. For the mixed $\trip$, we compute the solar capture rate $C$ using the inelastic cross section in Eq.~\eqref{eq:HXM_sigma32}, while obtaining the capture--annihilation equilibration time by rescaling the $\mathbf{3}_1$ result of the same reference.

In our mixed scenario, the neutral states predominantly originate from the Dirac pair formed by the two $\mathbf{3}_0$ multiplets. We therefore assess the differences in annihilation and elastic scattering between the $\mathbf{3}_0$ and $\mathbf{3}_1$.
We retain the reference annihilation cross section, since the total EW annihilation rates of the two multiplets differ only by small hypercharge-dependent corrections within the
off-resonance Sommerfeld approximation adopted in Ref.~\cite{Luengas:2026tlf}. For the masses relevant to our bound, loop-induced spin-independent elastic scattering is sufficient to thermalize the captured $\mathbf{3}_1$ population over the solar lifetime. Thermalization is expected to be at least as efficient for the predominantly $\mathbf{3}_0$ state, whose elastic cross section is larger~\cite{Bloch:2024suj,Bottaro:2022one}. Both populations thus reach the same thermal distribution at fixed DM mass, allowing us to adopt the same effective annihilation volume.

Under these approximations, the equilibration time changes only through the capture rate. We therefore rescale the cyan curve in Fig.~4 of Ref.~\cite{Luengas:2026tlf} as
\begin{equation}
\frac{\tau_{\rm eq}}{t_\odot}
\simeq
\left(\frac{\tau_{\rm eq}}{t_\odot}\right)_{\rm ref}
\sqrt{\frac{C_{\rm ref}}{C}},
\end{equation}
where the reference quantities are evaluated at the same DM
mass and at the splitting satisfying
$C_{\rm ref}/2=\Gamma_{WW}^{\rm lim}$, as prescribed for that figure. Here $t_\odot$ denotes the age of the Sun. Using our calculated $C$ and the rescaled $\tau_{\rm eq}$, we obtain the bound by comparing the annihilation rate
\begin{equation}
\Gamma_{WW}^{\rm lim}\simeq
\frac{C}{2}\tanh^2\left(\frac{t_\odot}{\tau_{\rm eq}}\right)
\end{equation}
with the IceCube limit, taking $W^+W^-$ to dominate annihilation.

\end{document}